\PassOptionsToPackage{varg}{txfonts}
\documentclass{aa}  
\usepackage{graphicx}
\usepackage{txfonts}
\usepackage{lipsum}
\usepackage{subcaption}         
\usepackage{lscape}             
\usepackage{placeins}           

\usepackage[switch]{lineno}
\usepackage{caption}
\usepackage{float}
\usepackage{tabularx}
\usepackage{array}
\usepackage{CJKutf8}
\usepackage{siunitx}
\usepackage{afterpage}
\usepackage{CJK}
\usepackage{silence}
\usepackage{ulem}
\usepackage{placeins}
\usepackage{booktabs}
\newcommand{\ha} {\mbox{H$\alpha$}\,}
\newcommand{\hb} {\mbox{H$\beta$}\,}
\newcommand{\hg} {\mbox{H$\gamma$}\,}

\newcommand{\Naid}{\ion{Na}{id}\,}
\newcommand{\Feii} {\ion{Fe}{ii}\,}

\newcommand{\Caii} {\ion{Ca}{ii}\,}

\newcommand{\Oi} {\ion{O}{i}\,}

\newcommand{\Scii} {\ion{Sc}{ii}\,}

\newcommand{\Baii} {\ion{Ba}{ii}\,}

\usepackage{etoolbox} 
\makeatletter
\patchcmd{\AA@hyperref}
  {\AddToHook{begindocument}{showkeys}{before}{nameref}}
  {}{}{}
\makeatother

\usepackage{hyperref}

\begin{document}

\title{A prolonged plateau-to-tail transition in the Type II supernova SN\,2025abyc}

\titlerunning{SN~2025abyc}

\author{
Luhan Li \inst{1,2,3}\corrauth{liluhan@ynao.ac.cn} 
\and Bo Wang \inst{1,2,3}\corrauth{wangbo@ynao.ac.cn}
\and Jujia Zhang \inst{1,2} \email{jujia@ynao.ac.cn}
\and Zhengyang Zhang \inst{1,2,3}\email{zhangzhengyang@ynao.ac.cn}
\and Xinjie Luo \inst{1,2,3}\email{luoxinjie@ynao.ac.cn}
\and Shiyang Dong \inst{1,2,3}\email{dongshiyang@ynao.ac.cn}
\and Saien Xu  \inst{4}\email{saienxu@whu.edu.cn}
\and Zhengwei Liu \inst{1,2}\email{zwliu@ynao.ac.cn}
\and Zhanwen Han \inst{1,2}\email{zhanwenhan@ynao.ac.cn}
}

\institute{
Yunnan Observatories, Chinese Academy of Sciences, Kunming 650216, China
\and International Centre of Supernovae (ICESUN), Yunnan Key Laboratory of Supernova Research, Kunming 650216, China
\and University of Chinese Academy of Sciences, Beijing 100049, China
\and Department of Astronomy, School of Physics and Technology, Wuhan University, Wuhan 430072, China
}

 
\abstract{
We present optical photometric and spectroscopic observations of the Type II supernova (SN) SN\,2025abyc.
During the optically thick phase between approximately 10 and 70\,d after explosion, its light curves show strongly wavelength-dependent decline rates of approximately 2.7, 2.1, 0.9, and 0.8\,mag\,(100\,d)$^{-1}$ in the $g$, $c$, $r$, and $o$ bands, respectively.
At approximately 70\,d, the light curves begin to depart from their nearly linear plateau evolution and gradually transition toward the radioactive tail. 
A Fermi--Dirac fit to the well-sampled ATLAS $o$-band light curve yields a transition midpoint of $t_{\rm PT}\approx100.5$\,d. 
The interval between the end of the linear plateau and this transition midpoint is approximately 30\,d, indicating a prolonged plateau-to-tail transition. 
This timescale is comparable to those measured for SN\,2013by, SN\,2013ej, and SN\,2014G.
Spectroscopically, at +13\,d post-explosion, the \ha\ profile appears weak and broad, whereas \hb\ and \hg\ display clear P-Cygni profiles. 
This morphology can be explained by the normal early spectroscopic evolution of SNe II, although partial filling of the \ha\ absorption trough by emission associated with circumstellar interaction cannot be excluded.
SN\,2025abyc otherwise follows the general photospheric velocity evolution of SNe II, while remaining toward the high-velocity side of the comparison distribution in \ha, \hb, and \Feii.
Exploratory light-curve modelling suggests a synthesized $^{56}$Ni mass of approximately $0.03$--$0.04\,M_\odot$. 
We suggest that the extended circumstellar environment, $^{56}$Ni distribution, and hydrogen-envelope structure could all play a role in shaping the observed light-curve evolution, particularly the prolonged plateau-to-tail transition.
}

\keywords{supernovae: general -- supernovae: individual: SN 2025abyc -- stars: circumstellar matter}

\maketitle
\nolinenumbers
 
\section{Introduction} \label{sect:intro}
Core-collapse (CC) supernovae (SNe) mark the terminal explosions of massive stars with initial masses $\gtrsim 8\,\mathrm{M_\odot}$, triggered by the gravitational collapse of their stellar cores \citep[see, e.g.,][]{2003ApJ...591..288H,2009ARA&A..47...63S}.
Among them, Type II supernovae (SNe II) are characterized by hydrogen features in their spectra and have historically been divided into several subclasses, including SNe IIP, SNe IIL, SNe IIn, and SNe IIb \citep[e.g.,][]{1997ARA&A..35..309F}.
However, the traditional photometric distinction between SNe IIP and IIL has been challenged by large-sample studies. 
Their light-curve decline rates and spectroscopic properties show continuous distributions, without a clear boundary separating the two subclasses \citep[e.g.,][]{2014ApJ...786...67A,2015ApJ...799..208S,2015MNRAS.448.2608V,2017ApJ...850...90G}.
SNe IIP and IIL are therefore increasingly regarded as the slowly and rapidly declining ends of a continuous population of normal hydrogen-rich SNe II.

SNe IIP account for about 70\% of all nearby SNe II \citep[][]{2025A&A...698A.305M}.
These events exhibit a prominent, long-lasting plateau phase in multi-band light curves, typically spanning $\sim$60--140 days, particularly at redder wavelengths \citep[e.g.,][]{2014ApJ...786...67A,2024A&A...692A..95A}
.
This plateau is powered by the combined effects of thermalized shock-deposited energy and hydrogen recombination in the expanding envelope \citep[e.g.,][]{1993ApJ...414..712P,2026ApJ..1002...68L}.
Progenitors of SNe IIP have been observationally confirmed to be red supergiants (RSGs) through detections in pre-explosion images 
\citep[e.g.,][]{2009ARA&A..47...63S,2025Galax..13...33V,2025ApJ...982L..55L}.
To date, about 20 such progenitors have been directly identified \citep[see][]{2025Galax..13...33V}.

With the rapid development of wide-field transient surveys and advances in follow-up facilities, the number of discovered nearby SNe II has increased significantly.
Along with this growing sample, more events with diverse and peculiar observational properties have been identified.
Short-plateau SNe II have previously been reported, including SN\,2006Y, SN\,2006ai, and SN\,2016egz \citep[][]{2021ApJ...913...55H}, SN\,2020jfo \citep[][]{2022ApJ...930...34T}, SN\,2018gj \citep{2023ApJ...954..155T}, and SN\,2021tsz \citep{2025A&A...703A.224D}.
These events are generally interpreted as explosions of progenitors with relatively low-mass hydrogen-rich envelopes, potentially resulting from enhanced mass loss or binary interaction. 
Among them, SN 2018gj is particularly informative because its reported pre-explosion progenitor counterpart and old local stellar environment provide constraints on its evolutionary origin beyond its short plateau duration.
Based on these properties, \citet{2026arXiv260106577N} proposed that its progenitor was a binary-merger product, providing a possible observational link between short-plateau SNe II and progenitors formed through binary mergers.
In parallel, SN\,2023ufx is an extreme SN IIP characterized by an unusually short plateau ($\sim$47\,d) and rapid early decline, explained by a massive progenitor with a heavily stripped hydrogen envelope ($\sim$1.2\,$\mathrm{M_\odot}$) and high $^{56}$Ni production \citep[$\sim$0.14\,$\mathrm{M_\odot}$;][]{2025ApJ...982...12R}.
Binary-evolution calculations have shown that binary interactions can substantially modify the hydrogen-envelope mass and pre-SN structure, thereby producing a wider range of SN II progenitor and explosion properties \citep[e.g.,][]{2019A&A...631A...5Z,2021A&A...645A...6Z,2024A&A...685A.169D}.

Here we present SN\,2025abyc, an SN II whose optical light curves exhibit a prolonged plateau-to-tail transition. 
Following the end of the approximately linear plateau at $\sim70$\,d after explosion, the light curves enter the plateau-to-tail transition.
The interval between the departure from the linear plateau and the fitted transition midpoint at $t_{\rm PT}\approx100.5$\,d is approximately 30\,d, indicating a morphologically prolonged transition.
This extended interval of progressive steepening produces a morphologically prolonged plateau-to-tail evolution relative to the comparison SNe II.
In Section~\ref{sect:basic_inf}, we report the basic target information and data reduction procedures.
The photometric and spectroscopic analyses of SN\,2025abyc are presented in Sections~\ref{sect:photometry} and~\ref{sect:spectroscopy}, respectively.
The discussion is given in Section~\ref{sect:discussion}, and conclusions are presented in Section~\ref{sect:conclusion}.
Relevant data tables are provided in the Appendix.

\section{Basic information and data reduction} \label{sect:basic_inf}
\subsection{Explosion epoch, extinction, and distance}
\begin{figure}
   \centering
   \includegraphics[width = 0.95\linewidth]{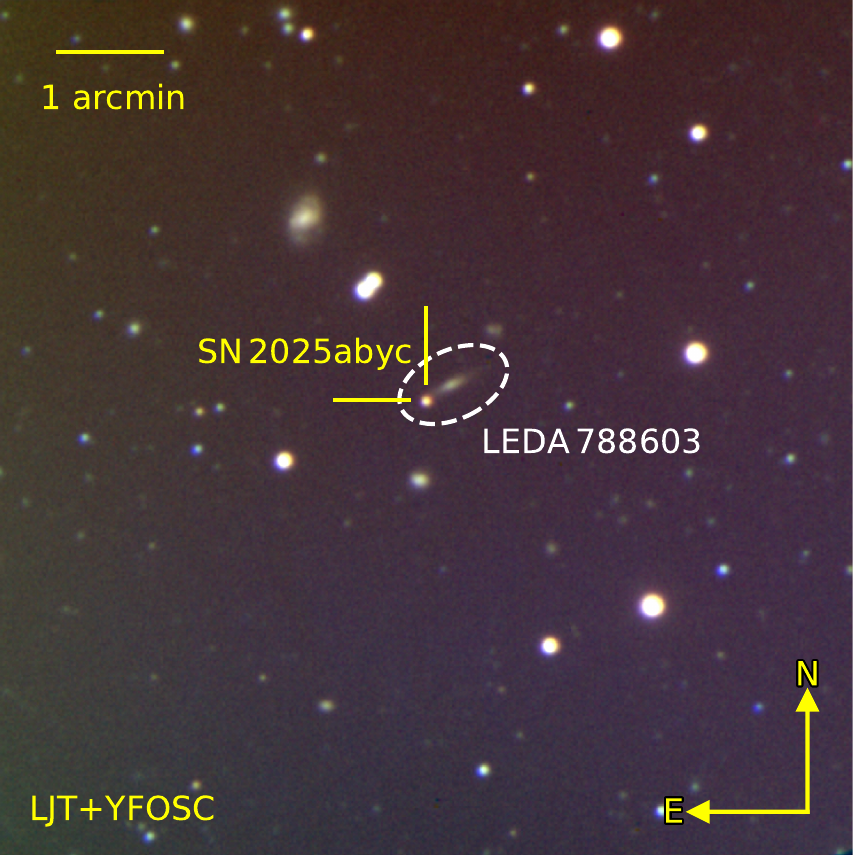}
      \caption{Location of SN\,2025abyc in a combined $gri$ image obtained with the LJT by stacking observations from multiple epochs. 
      The image orientation and scale are indicated.}
     \label{Fig:location}
\end{figure}

\begin{table}
\centering
\caption{Basic properties and adopted parameters of SN\,2025abyc}
\label{tab:properties}
\renewcommand{\arraystretch}{1.1}
\begin{tabular}{@{}lr@{}} 
\hline\hline
RA (J2000) & 02:44:25.313\\
DEC (J2000) & -24:34:08.62\\
Host Galaxy & LEDA~788603\\
Redshift $z$ & $0.021\pm0.002$\\
$E(B - V)_{\mathrm{Gal}}$ & $0.02 \, \mathrm{mag}$\\
$E(B - V)_{\mathrm{host}}$ & $\sim 0 \, \mathrm{mag}$\\
Estimated Explosion Epoch (MJD)$^{\mathrm{1}}$ & $60974.62^{+0.02}_{-0.02}$\\
SCM distance (Mpc) & $82 \pm 9$ \\
\hline
\end{tabular}
\vspace{0.1mm}
\begin{flushleft}
Notes:
1. The quoted uncertainty includes only the statistical uncertainty from the fit.
\end{flushleft}
\end{table}

\begin{figure*} 
   \centering
    \sidecaption
   \includegraphics[width = 0.7\textwidth]{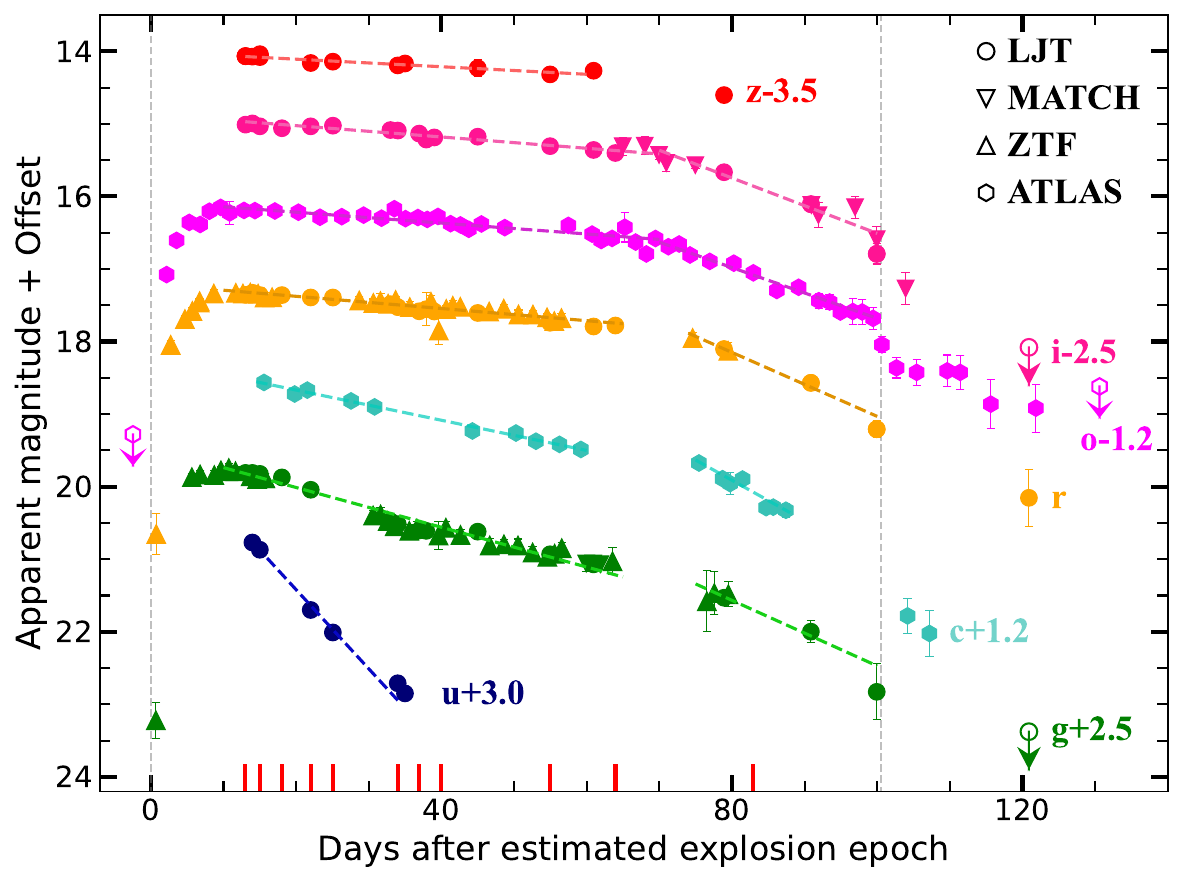}
   \caption{Multi-band apparent light curves of SN\,2025abyc obtained with the LJT, MATCH, ZTF, and ATLAS.
   The two vertical dashed gray lines represent the explosion epoch (phase zero) and the midpoint of the plateau-tail transition phase, respectively.
   Here, we adopted the explosion epoch $\rm MJD~=~60974.62$.
   We also overplot linear fits to illustrate the different decline rates.
   For clarity, all light curves have been shifted vertically by different offsets.
   The ATLAS $co$-band observations have been combined into daily stacks. 
   The epochs of spectroscopic observations are marked by vertical solid red lines at the bottom of the figure.
   Error bars are smaller than the symbol sizes in most cases.}
    \label{Fig:LC}
\end{figure*}  

SN\,2025abyc was discovered by \citet{2025TNSTR4358....1T} on 28.53 October 2025 (MJD~=~60976.53; all dates are in UT unless noted otherwise), with an $o$-band magnitude of $18.47\,\mathrm{mag}$.
The last non-detection was obtained on 24.18 October 2025 (MJD~=~60972.18) at $o > 19.27\,\mathrm{mag}$.
It was classified as an SN II by the University of Hawaii 88-inch telescope \citep[UH88;][]{2025TNSCR4452....1J} from a spectrum taken on 2 November 2025 (MJD~=~60981.45).
The J2000 coordinates are $\mathrm{RA}=02^{\mathrm h}44^{\mathrm m}25.313^{\mathrm s}$ and $\mathrm{Dec}=-24^{\circ}34^{\prime}08.62^{\prime\prime}$.
Its likely host is LEDA~788603 (also named WISEA J024424.29-243359.3), located $15.26^{\prime\prime}$ west and $9.29^{\prime\prime}$ north of the SN position.
The SN location within the host galaxy is shown in Figure~\ref{Fig:location}, and relevant property parameters are summarized in Table~\ref{tab:properties}.
It is worth noting that the recently updated NASA/IPAC Extragalactic Database Local Volume Sample (NED-LVS) reported a nearby object, APMUKS(BJ) B024210.33$-$244637.1, at the position of the host galaxy \citep{2023ApJS..268...14C,NED8}. 
This source is separated by only $\sim4.3^{\prime\prime}$ from LEDA~788603.
No redshift measurement for this source is available in major public databases, including NED and HyperLEDA \citep{1991ASSL..171...89H,2003A&A...412...45P}.

To better constrain the explosion epoch, we used early-time photometry from the Zwicky Transient Facility \citep[ZTF;][]{2019PASP..131a8003M} and the Asteroid Terrestrial-impact Last Alert System \citep[ATLAS;][]{2018PASP..130f4505T,2020PASP..132h5002S}.
The two earliest significant ZTF detections, at MJD$\sim$60975.3, were obtained from forced photometry retrieved through the ALeRCE broker \citep{2021AJ....161..242F}, and have photometric uncertainties of approximately 0.2\,mag.
Several earlier ZTF forced-photometry measurements have uncertainties of approximately 0.6\,mag and are therefore treated as low-significance flux measurements rather than detections.
We retained these measurements with their corresponding uncertainties in the fit, because forced photometry can recover low-significance pre-discovery flux and provide improved constraints on the first-light epoch \citep{2025JCAP...08..053G}.
The subsequent ZTF measurements are standard alert-stream detections.
The ATLAS $o$-band forced photometry was retrieved from the ATLAS forced-photometry server \citep{2018PASP..130f4505T,2020PASP..132h5002S}, and same-filter exposures obtained during the same night were combined into daily stacks using sigma clipping for outlier rejection.
We fitted the early-time ATLAS and ZTF flux measurements with an expanding fireball model, $F(t)=F_1\,(t-t_0)^2$.
The fit is shown in Figure~\ref{Fig:explosion_epoch}, and yields $t_0=\mathrm{MJD}\,60974.62^{+0.02}_{-0.02}$, which is adopted throughout this work.
The quoted uncertainty represents the statistical uncertainty from the fireball-model fit only.

For interstellar reddening, we adopt $E(B-V)_{\rm Gal}=0.02$\,mag from NED \citep{Schlafly2011ApJ...737..103S}, assuming a reddening law with $R_V=3.1$ \citep{Cardelli1989ApJ...345..245C}.
We do not identify a statistically significant narrow interstellar \Naid\ absorption component at the host-galaxy redshift in the
available early-time spectra.   
However, given the spectral resolution and
signal-to-noise ratio (S/N), this non-detection does not provide a stringent upper limit on the host extinction.
Together with the projected SN position in the host outskirts, this suggests that host-galaxy extinction is likely small.
We therefore adopt $E(B-V)_{\rm total}=0.02$\,mag.

\citet{2025TNSCR4452....1J} reported an initial estimate of $z\sim0.02$ from the first classification spectrum.
Because no narrow host-galaxy emission lines are detected, we estimated the SN rest-frame redshift using broad-feature alignment in the early Lijiang 2.4 m telescope (LJT) spectrum and the Next Generation SuperFit (NGSF) template matching \citep{2022TNSAN.191....1G,2005ApJ...634.1190H}.
The best-fitting templates suggest $z=0.018$--0.023 (with good matches to SN\,2012aw and SN\,2007od), and we adopt $z=0.021\pm0.002$ for SN\,2025abyc in this work.

Given the lack of a direct host recession-velocity measurement, we estimated the distance using the standard candle method (SCM) for SNe II \citep[e.g.,][]{2002ApJ...566L..63H}, based on the empirical luminosity--velocity relation.
For SN\,2025abyc, we derived the apparent Johnson--Cousins $V$-band magnitude at 50\,d post-explosion by interpolating $g$- and $r$-band photometry at $+48.6$\,d and $+50.6$\,d, and transforming $(g,r)$ to $V$ using the Sloan Digital Sky Survey (SDSS) colour relations of \citet{Lupton2005SDSSTransform}.
The Fe II $\lambda5169$ velocity at 50\,d was obtained by interpolating between spectra taken at $+40.0$\,d and $+55.0$\,d.
Using $m_V^{50}=17.89\pm0.05$\,mag, $A_V=0.062\,$mag, and $v_{\rm FeII}=4600\pm390\,\mathrm{km\,s^{-1}}$, we obtain an SCM distance of $82 \pm 9$\,Mpc based on the calibration of \citet{2002ApJ...566L..63H}.
This is consistent within uncertainties with a Hubble-flow distance of $\sim86$\,Mpc for $z=0.021$ assuming a standard cosmology with $H_0=73\pm5\,\mathrm{km\,s^{-1}\,Mpc^{-1}}$ \citep{Spergel2007ApJS..170..377S}.
We adopt the SCM distance in the following analysis.

\subsection{Data reduction}

Optical photometry of SN\,2025abyc in the \textit{ugriz} bands was obtained with the LJT equipped with the Yunnan Faint Object Spectrograph and Camera \citep[LJT+YFOSC;][]{2015RAA....15..918F,2019RAA....19..149W} over the modified Julian date (MJD) range 60987--61095.
Additional observations in the \textit{g} and \textit{i} bands were carried out with the Multi-mode Autonomous Terminal for Compatible Hybrid-optics (MATCH), operated by Wuhan University. 
Aperture photometry for the LJT images was performed using the AutoPhOT pipeline \citep{2022A&A...667A..62B}.
For images that could not be processed by AutoPhOT, we performed aperture photometry using comparison stars with standard \texttt{IRAF} routines \citep{Tody1986SPIE..627..733T,Tody1993ASPC...52..173T}.
The resulting LJT and MATCH photometric measurements are listed in Table~\ref{tab:phot_data} in the Appendix.

The ZTF began monitoring SN\,2025abyc (ZTF25acbcfzr) on 27 October 2025 (MJD~=~60975) in the \textit{g} and \textit{r} bands.
To complement our dataset, we downloaded the public ZTF data\footnote{\url{https://alerce.online/object/ZTF25acbcfzr}} from first detection to 21 January 2026 \citep[MJD~=~61061;][]{2019PASP..131a8003M,2021AJ....161..242F}.
We also collected ATLAS $c$- and $o$-band photometry from the ATLAS survey \citep{2018PASP..130f4505T,2020PASP..132h5002S}.
The forced-photometry light curves were retrieved from the ATLAS data-release server\footnote{\url{https://fallingstar-data.com/forcedphot/}} \citep{Shingles2021TNSAN...7....1S}.
To improve data quality, we used the public stacking tool\footnote{\url{https://github.com/thespacedoctor/plot-results-from-atlas-force-photometry-service}} to combine same-filter measurements into daily stacks with sigma clipping for outlier rejection.

Spectroscopic observations of SN\,2025abyc were obtained with LJT/YFOSC \citep{2019RAA....19..149W}, and the observing log is provided in Table~\ref{table:specinfo} in the Appendix.
The data were reduced using standard \texttt{IRAF} procedures, including bias and flat-field corrections, wavelength calibration, atmospheric-extinction correction, and flux calibration with spectrophotometric standard stars observed at similar airmasses.
Telluric absorption was corrected using spectrophotometric standard stars.
We extracted the 6700--8400\,\AA\ region of each standard-star spectrum using the \texttt{imcopy} task in \texttt{IRAF} \citep{Tody1986SPIE..627..733T,Tody1993ASPC...52..173T} and continuum-normalized the extracted spectrum to construct a telluric-transmission template.
The template was then shifted and scaled to match the telluric absorption in each SN spectrum, and the correction was applied using the \texttt{telluric} task.
In addition, an early-time public spectrum obtained with the 2.2\,m University of Hawaii telescope (UH88) was retrieved from the Weizmann Interactive Supernova Data Repository \citep[WISeREP;][]{2012PASP..124..668Y} to improve early spectral-phase coverage.

\section{Photometry} \label{sect:photometry}

\begin{figure}
    \centering
    \includegraphics[width=0.98\linewidth]{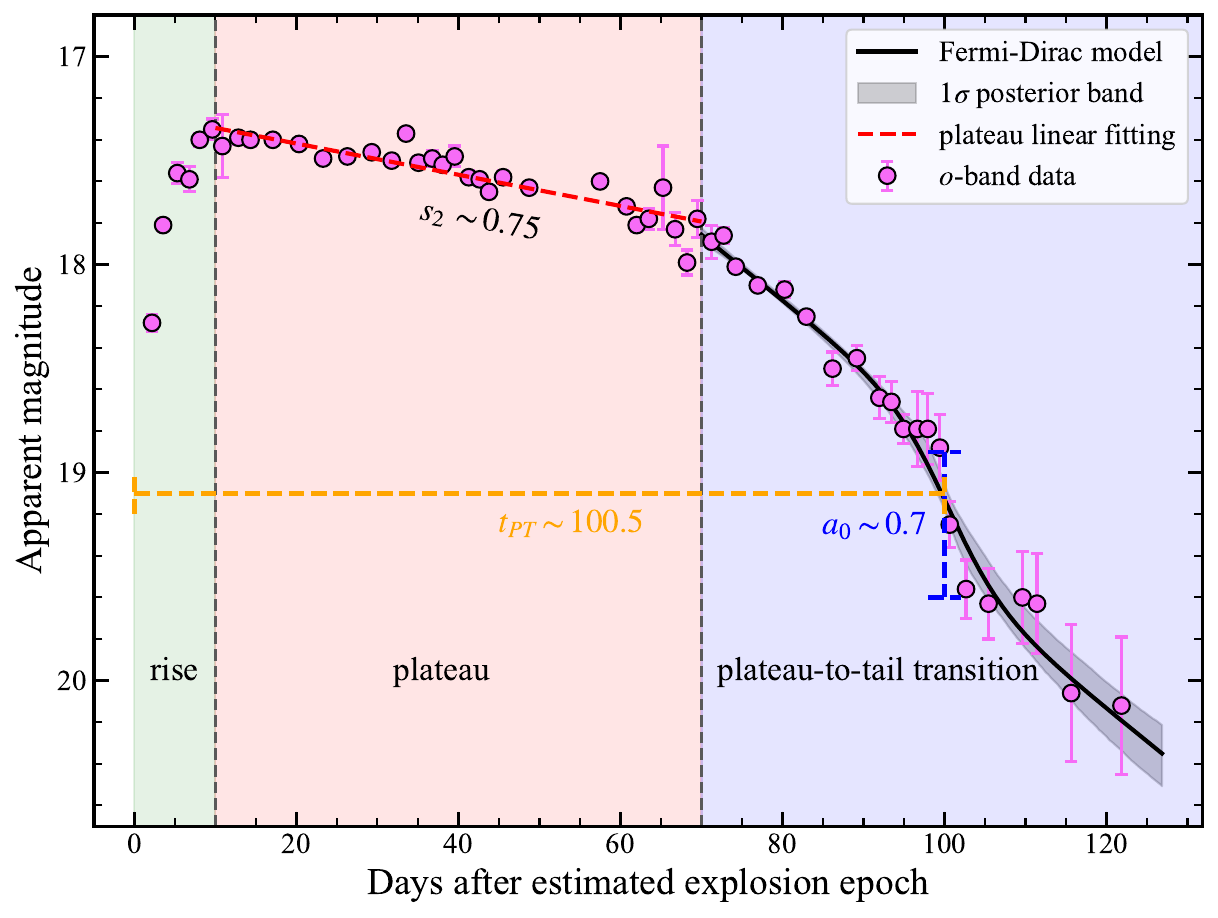}
    \caption{The apparent $o$-band light curve of SN\,2025abyc. 
    For the purpose of characterizing the light-curve evolution, we divide it into three phases: the rise phase (0--10 d), the plateau phase (10--70 d), and the plateau-to-tail transition phase (after 70 d).
    The solid black curve shows the best-fitting Fermi--Dirac function to the transition.}
    \label{fig:phase}
\end{figure} 

\begin{table}
    \centering
    \caption{Multi-band light-curve decline rates of SN\,2025abyc at different phases.}
    \renewcommand{\arraystretch}{1.0}
    \setlength{\tabcolsep}{5pt}
    \begin{tabular}{cccc}
        \hline \hline
        Band & $t_{\rm start} (\mathrm{d}) $ & $t_{\rm stop} (\mathrm{d})$ & Slope ($\rm mag\, /100\,d$)\\
        \hline
        u & 10 & 70 &10.95$\pm$0.14 \\
        \hline
        g & 10 & 70 &2.74$\pm$0.06 \\
        g & 70 & 100 &4.52$\pm$1.10 \\
        \hline        
        c & 10 & 70 &2.08$\pm$0.05 \\
        c & 70 & 100 &5.58$\pm$0.60 \\
        \hline
        r & 10 & 70 & $0.87\pm0.04 $\\
        r & 70 & 100 & $4.41\pm0.41 $\\
        \hline
        o & 10 & 70 &0.75$\pm$0.04\\
        o & 70 & 100 &3.50$\pm$0.19\\
        o & 100 & 120 &3.93$\pm$1.18\\
        \hline
        i & 10 & 70 & 0.78$\pm$0.06\\
        i & 70 & 100 &3.82$\pm$0.27\\
        \hline
        z & 10 & 70 &0.52$\pm$0.12\\
        \hline
    \end{tabular}
    \label{tab:decline_rate}
\end{table}

\begin{figure}
    \centering
    \includegraphics[width=0.98\linewidth]{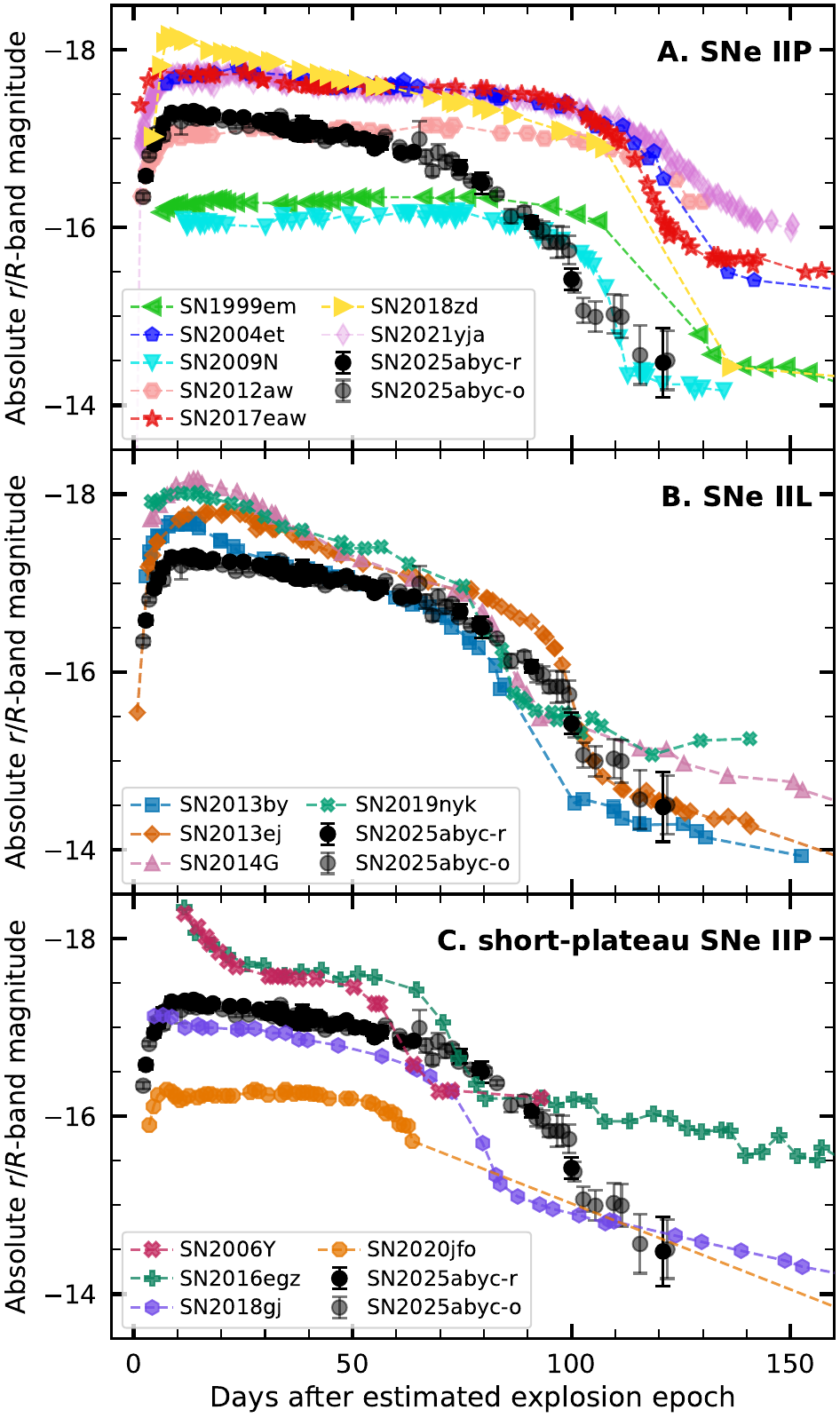}
    \caption{The absolute $r/R$-band light curves of SN\,2025abyc compared to those of well-observed SNe II.
    The three panels present comparisons with different types of Type II SNe: normal SNe IIP (panel A), fast-declining SNe IIL (panel B), and short-plateau SNe IIP (panel C).
    The ATLAS absolute $o$-band data of SN\,2025abyc are also included (grey circles) to enhance the temporal coverage of the light curve.
    }
    \label{fig:Abs_r}
\end{figure}

\subsection{Apparent light curves} \label{sect:Apparent_LC}

Our optical follow-up observations of SN\,2025abyc began on MJD$\sim$60987 and continued until MJD$\sim$61095 ($+120$\,d after explosion).
The monitoring was conducted in the $ugriz$ bands with the LJT and in the $gi$ bands with the MATCH, supplemented by public ZTF $gr$-band and ATLAS $co$-band photometry.
The multi-band apparent light curves are shown in   Figure~\ref{Fig:LC}.
As described in Section~\ref{sect:basic_inf}, ATLAS $c$- and $o$-band measurements were combined into daily stacks to improve the S/N.
The rise is well sampled in the $g$, $r$, and $o$ bands.
To estimate the peak magnitudes during the rise, we fitted the early-time light curves in each band independently with a second-order polynomial using a Markov chain Monte Carlo (MCMC) approach based on \texttt{emcee} package in Python \citep{2013PASP..125..306F}.
We find a $g$-band peak of $17.17\pm0.03$\,mag at $\mathrm{MJD}=60985.2\pm0.2$, corresponding to a rise time of $10.6\pm0.2$\,d.
For the $r$ band, the peak is $17.29\pm0.02$\,mag at $\mathrm{MJD}=60986.3\pm0.5$, with a rise time of $11.7\pm0.5$\,d.
In the ATLAS $o$ band, the peak is $17.31\pm0.02$\,mag at $\mathrm{MJD}=60985.5\pm0.3$, corresponding to a rise time of $10.9\pm0.3$\,d.

After the initial rise, the multi-band light curves enter the optically thick plateau phase.
To illustrate the different stages of the light-curve evolution, Figure~\ref{fig:phase} presents the well-sampled ATLAS $o$-band light curve.
Based on visual inspection of its morphology, we divide the evolution into an initial rise at $\lesssim10$\,d, a plateau phase between $\sim10$ and 70\,d, a plateau-to-tail transition beginning at approximately 70\,d, followed by the radioactive tail.
These approximate boundaries are intended as descriptive guides rather than statistically determined change points.
For a descriptive comparison of the decline rates, we performed linear regressions over two intervals, $\sim10$--70\,d and $\sim70$--100\,d,
in each available band. 
The first interval here corresponds to the plateau phase \citep[$s_2$,][]{2014ApJ...786...67A}, whereas the second samples the early part of the plateau-to-tail transition.
The resulting decline rates for all bands are listed in Table~\ref{tab:decline_rate} and overplotted in Figure~\ref{Fig:LC}.
In most bands, the earlier interval (10--70\,d) shows a relatively shallow decline, with rates of $\sim$0--3\,mag\,(100\,d)$^{-1}$, whereas the later interval (70--100\,d) is significantly steeper, with rates of $\sim$4--5\,mag\,(100\,d)$^{-1}$.

To characterise the plateau-to-tail evolution more quantitatively, we fitted the ATLAS $o$-band light curve from the plateau through the transition and into the radioactive tail using a Fermi--Dirac function
\citep[see, e.g.,][]{2010ApJ...715..833O,2016MNRAS.459.3939V,2021ApJ...906...56D,2021MNRAS.505.1742R,2025ApJ...982...12R}:
\begin{equation}
m(t)=-\frac{a_0}{1+\exp[(t-t_{\rm PT})/w_0]}+p_0(t-t_{\rm PT})+m_0,
\end{equation}
where $m(t)$ is the apparent magnitude at phase $t$, measured relative to the explosion epoch.
The parameter $a_0$ represents the asymptotic magnitude change associated with the plateau-to-tail transition after accounting for the underlying linear evolution.
The parameter $t_{\rm PT}$ denotes the midpoint of the transition, at which half of the modelled transition amplitude has been reached.
The parameter $w_0$ is the characteristic width of the transition and controls its sharpness, with larger values corresponding to a more gradual transition.
The coefficient $p_0$ describes the linear decline rate of the radioactive tail, while $m_0$ is the magnitude of the underlying linear component at $t=t_{\rm PT}$.

The best-fitting model is shown in Figure~\ref{fig:phase}, while the posterior distributions of the fitted parameters are presented in
Figure~\ref{Fig:FDcorner}.
The fit yields $t_{\rm PT}\approx100.5$\,d and fitted amplitude $a_0=0.7^{+0.5}_{-0.2}$\,mag.
Compared with the $a_0$ range of $0.5$--$3.5$\,mag reported by \citet{2016MNRAS.459.3939V}, SN\,2025abyc lies close to the lower boundary of this distribution.
The fitted characteristic transition width is $w_0=3.4^{+3.9}_{-2.4}$\,d.
The inferred $w_0$ lies within the typical range of approximately 2--5\,d found for SNe II \citep[][]{2016MNRAS.459.3939V}.
Thus, neither $a_0$ nor $w_0$ is individually a clear outlier relative to previous SN II samples.
Nevertheless, the observed light curve of SN\,2025abyc shows a morphologically prolonged evolution toward the radioactive tail.
It begins to depart from its approximately linear plateau decline at around 70\,d, approximately 30\,d before the fitted transition midpoint at $t_{\rm PT}\approx100.5$\,d.
We stress that this morphological interval is not equivalent to the formal Fermi--Dirac width $w_0$.

\subsection{Absolute light curve}

Figure~\ref{fig:Abs_r} compares the optical $r$-band light curves of SN\,2025abyc with a subset of well-observed SNe II, divided into three categories: normal SNe IIP, fast-declining SNe IIL, and short-plateau SNe IIP.
The adopted distances, redshifts, and reddening values of these comparison objects from the literature are summarized in Table~\ref{tab:SNe_II_info} in the Appendix.
The comparison objects were selected based on the availability of well-sampled photometric and spectroscopic observations and reliable explosion parameter estimates.
For some comparison SNe without available $r$-band observations, we adopt the photometry in the adjacent $R$ band as a proxy.
For SN\,2025abyc, we show the absolute magnitude in the ATLAS $o$ band (grey points) in addition to the $r$-band light curve to extend temporal coverage.

From our photometry, SN\,2025abyc reaches a peak absolute magnitude of $M_r=-17.34\pm0.25$ in the observed $r$ band (distance and extinction as in Section~\ref{sect:basic_inf}), where the uncertainty includes both the fitting uncertainty and the distance uncertainty.
This is slightly fainter than several luminous SNe IIP (see Panel A), including SN\,2004et \citep[$M_R=-17.9$\,mag;][]{2006MNRAS.372.1315S}, SN\,2017eaw \citep[$M_R=-17.9$\,mag;][]{2019ApJ...876...19S}, SN\,2021yja \citep[$M_R=-17.8$\,mag;][]{2022ApJ...935...31H}, and SN\,2018zd \citep[$M_r=-18.2$\,mag;][]{2020MNRAS.498...84Z}.
We note that the distances and reddening values adopted for the comparison SNe were compiled from different literature sources and were derived using heterogeneous methods.
Consequently, residual systematic uncertainties may remain in their absolute magnitudes.
Using a sample of 330 ZTF SNe IIP, \citet{2025PASP..137d4203D} reported that the $r$-band peak absolute magnitudes span roughly $-14$ to $-19$\,mag, with a luminosity function peaking near $-17$\,mag.
Given the uncertainty of the SCM distance adopted for SN\,2025abyc and the heterogeneous distance estimates available for the comparison objects, SN\,2025abyc is broadly consistent with the typical peak luminosity of SNe~IIP in the comparison sample.

In Figure~\ref{fig:Abs_r}, Panel B compares SN\,2025abyc with several well-studied SNe IIL, including SN\,2013by \citep{2015MNRAS.448.2608V}, SN\,2013ej \citep{2014MNRAS.438L.101V}, SN\,2014G \citep{2016MNRAS.455.2712B}, and SN\,2019nyk \citep{2024A&A...685A..44D}.
These events exhibit substantially steeper plateau declines than the canonical SNe IIP shown in Panel A.
To provide a consistent comparison, we performed linear fits to the $r/R$-band light curves over a common interval around 50\,d after explosion.
The resulting decline rate for SN\,2025abyc is approximately $1.0$\,mag\,(100\,d)$^{-1}$, consistent with the value of approximately $0.9$\,mag\,(100\,d)$^{-1}$ measured over the broader 10--70\,d interval in Table~\ref{tab:decline_rate}.
The corresponding decline rates are approximately $1.5$, $1.7$, $2.4$, and $1.3$\,mag\,(100\,d)$^{-1}$ for SN\,2013by, SN\,2013ej, SN\,2014G, and SN\,2019nyk, respectively.
Thus, the $r/R$-band plateau decline of SN\,2025abyc is shallower than those of the rapidly declining comparison events, but steeper than those of most canonical SNe IIP in Panel A, which generally show rates below $0.5$\,mag\,(100\,d)$^{-1}$.
An exception is SN\,2018zd, for which we measure a decline rate of approximately $1.1$\,mag\,(100\,d)$^{-1}$.

In the $g$ and ATLAS $c$ bands, SN\,2025abyc declines at approximately $2.7$ and $2.1$\,mag\,(100\,d)$^{-1}$, respectively.
\citet{2014MNRAS.445..554F} proposed that SNe II declining by more than $0.5$\,mag in the $V$ band between peak brightness and 50\,d after explosion can be classified photometrically as SNe IIL.
SN\,2025abyc fades by approximately $1.1$\,mag in the $g$ band over the corresponding interval and would therefore satisfy this criterion if it were applied directly to the $g$ band.
However, because the criterion was defined in the $V$ band, while the available measurements are in the $g$ and broad ATLAS $c$ bands, this comparison should be regarded as indicative rather than definitive.
Overall, the relatively rapid evolution of SN\,2025abyc in the blue bands (e.g., $g$ and $c$) is IIL-like, whereas its $r$-band decline lies between those of the canonical IIP and rapidly declining IIL comparison samples.

Panel C compares SN\,2025abyc with several short-plateau SNe IIP, including SN\,2006Y \citep{2021ApJ...913...55H}, SN\,2016egz \citep{2021ApJ...913...55H}, SN\,2018gj \citep{2023ApJ...954..155T}, and SN\,2020jfo \citep{Ailawadhi2023MNRAS.519..248A}.
For SN\,2025abyc, the approximately linear plateau evolution ends at around 70\,d, after which the light curve undergoes a gradual plateau-to-tail transition, as illustrated in Figure~\ref{fig:phase}.
The Fermi--Dirac fit yields $t_{\rm PT}\approx100.5$\,d, corresponding to the midpoint and approximately the maximum decline rate of the modelled transition.
Thus, approximately 30\,d elapse between the initial departure from the linear plateau evolution and the transition midpoint.
This extended interval of gradual steepening gives SN\,2025abyc a morphologically prolonged plateau-to-tail evolution.
In contrast, the short-plateau events in Panel C generally undergo more rapid transitions once they depart from their plateau evolution.
For example, the time interval between the end of the plateau\footnote{For clarity, we use the parameter OPTd, as defined by \citet{2014ApJ...786...67A}, to characterize the time corresponding to the end of the plateau. 
OPTd represents the time interval from the explosion epoch to the end of the optically thick phase and is similar to $t_{\rm plateau,end}$ in \citet{2026PASP..138b4204D}.
} and the midpoint of the transition, ($t_{\rm PT}-\mathrm{OPTd}$), is approximately 9\,d for SN\,2018gj \citep{2023ApJ...954..155T}, less than 7\,d for SN\,2020jfo \citep{2022ApJ...930...34T}, and about 10\,d for SN\,2016egz \citep{2021ApJ...913...55H}.
By comparison, we obtain intervals of approximately 27\,d, 28\,d, and 24\,d for SN\,2013by, SN\,2013ej, and SN\,2014G, respectively, using Fermi--Dirac fits to their $r/R$-band light curves and estimating the end of the plateau from the evolution of the time derivative of the light-curve slope, as shown in Figure~\ref{Fig:SNe_drop}.
These values are comparable to the approximately 30\,d interval measured for SN\,2025abyc.

\subsection{Colour evolution}
\begin{figure}
   \centering
   \includegraphics[width = 0.98\linewidth]{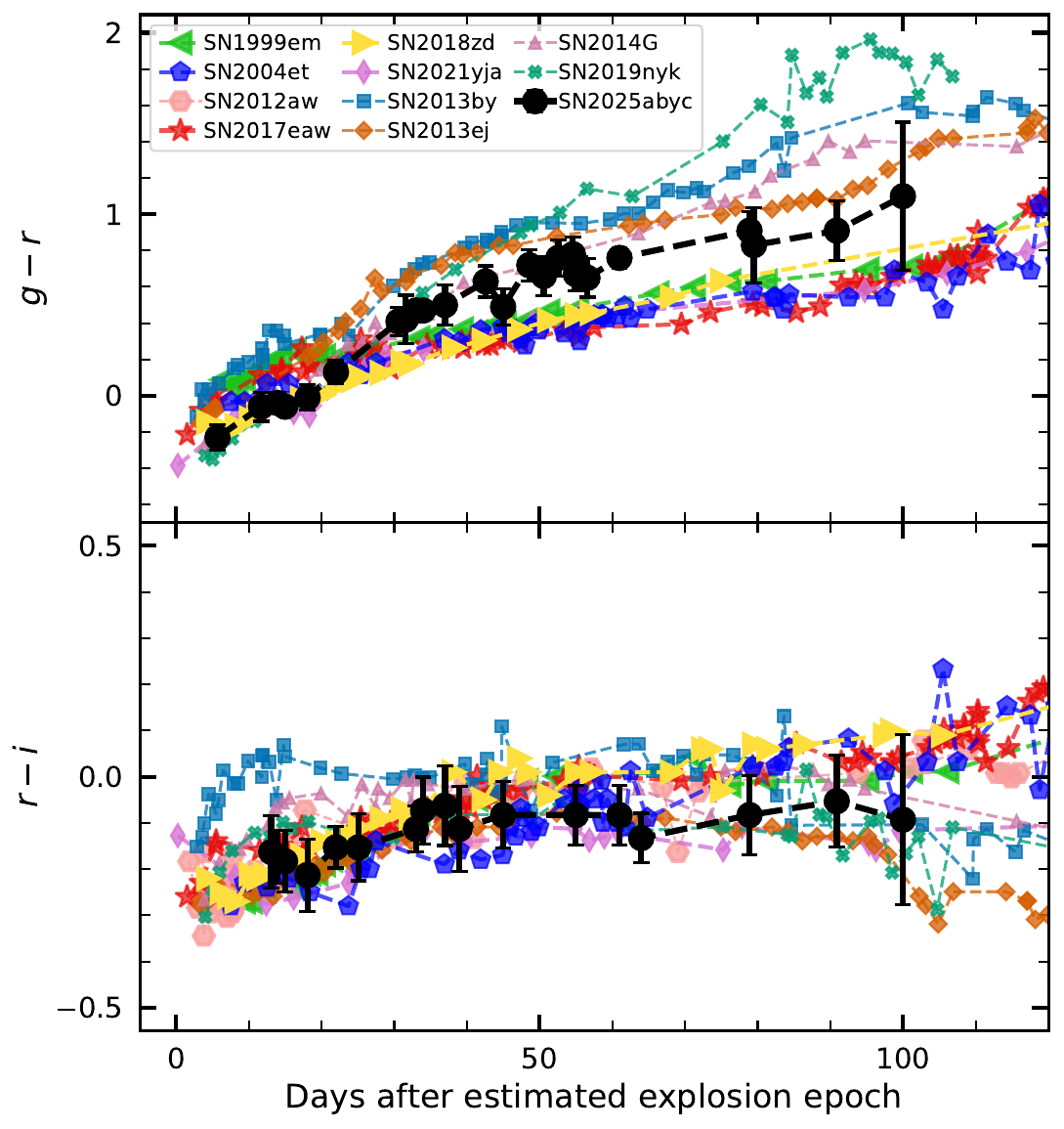}
   \caption{Colour evolution of SN\,2025abyc compared to other well-studied SNe II.
   All colours have been corrected for total extinction listed in Table~\ref{tab:SNe_II_info}.
   }
    \label{Fig:color}
\end{figure}  

Figure~\ref{Fig:color} presents the $(g-r)$ and $(r-i)$ colour evolution of SN\,2025abyc, compared with those of representative SNe IIP and IIL listed in Table~\ref{tab:SNe_II_info}.
For comparison objects without SDSS-like photometry, we converted Johnson--Cousins measurements using the colour transformations of \citet{2006A&A...460..339J}.
All colour curves have been corrected for the total line-of-sight extinction, including both Galactic and host-galaxy components, using the $E(B-V)_{\rm tot}$ values listed in Table~\ref{tab:SNe_II_info}.

As shown in Figure~\ref{Fig:color}, the SNe IIL events in our comparison sample, including SN\,2013by, SN\,2013ej, SN\,2014G, and SN\,2019nyk, generally evolve toward redder $(g-r)$ colours than the canonical SNe IIP.
The $(g-r)$ evolution of SN\,2025abyc lies approximately between the two groups, being generally redder than the canonical SNe IIP but bluer than most of the SNe IIL comparison objects.
This intermediate colour evolution is consistent with its photometric decline rates, which likewise place SN\,2025abyc between the canonical IIP and IIL events.
This tendency is qualitatively consistent with previous studies showing that faster-declining SNe II are, on average, slightly redder than more slowly declining events \citep{2014MNRAS.445..554F,2018MNRAS.476.4592D}.
By contrast, $(r-i)$ remains broadly consistent with the comparison samples over the same phase range.
This behaviour is qualitatively consistent with the multi-band light curves: the $g$ band is relatively brighter at early times but declines more rapidly at later phases than the $r$ and $i$ bands.
The spectral sequence supports this interpretation.
Early spectra are dominated by a hot, quasi-blackbody continuum with strong blue flux.
As the ejecta cool during the plateau phase, metal-line blanketing increases in the blue and \ha\ emission becomes more prominent relative to the continuum, producing the observed redward evolution in $(g-r)$.
This trend is consistent with a faster post-peak decline in the $g$ band than in most comparison SNe IIP, which would shift $(g-r)$ to redder values, while leaving $(r-i)$ broadly similar to that of the majority of the sample.

\citet{2018MNRAS.476.4592D} analysed the colour evolution of a sample of SNe II and found that their colour curves are generally characterised by two linear regimes.
The initial regime has a median slope of $s_{1,(g-r)}=1.90\pm0.53$\,mag\,(100\,d)$^{-1}$ and lasts until $37.9\pm4.3$\,d after explosion, while the subsequent regime has a shallower median slope of $s_{2,(g-r)}=0.68\pm0.25$\,mag\,(100\,d)$^{-1}$.
For comparison, we fitted the $(g-r)$ colour evolution of SN\,2025abyc with two linear segments separated at 38\,d, obtaining slopes of $2.64\pm0.11$ and $0.75\pm0.15$\,mag\,(100\,d)$^{-1}$ before and after the break, respectively.
The early-time slope lies above the median value, although it remains consistent with the dispersion of their sample.
This indicates that SN\,2025abyc undergoes somewhat more rapid early reddening, consistent with its faster colour evolution relative to the canonical SNe IIP shown in Figure~\ref{Fig:color}.
Such behaviour may reflect more rapid early photospheric cooling, although colour evolution alone does not uniquely constrain the ejecta expansion rate \citep[][]{1994A&A...282..731P,2018MNRAS.476.4592D}.
After 38\,d, the measured slope is slightly higher than, but fully consistent with, the median value of the comparison sample.
Overall, SN\,2025abyc exhibits the same two-regime colour evolution commonly observed in SNe II, with a somewhat faster evolution during the early phase.

\subsection{Bolometric light curve} \label{sect:Bol}

\begin{figure}
   \centering
   \includegraphics[width = 0.98\linewidth]{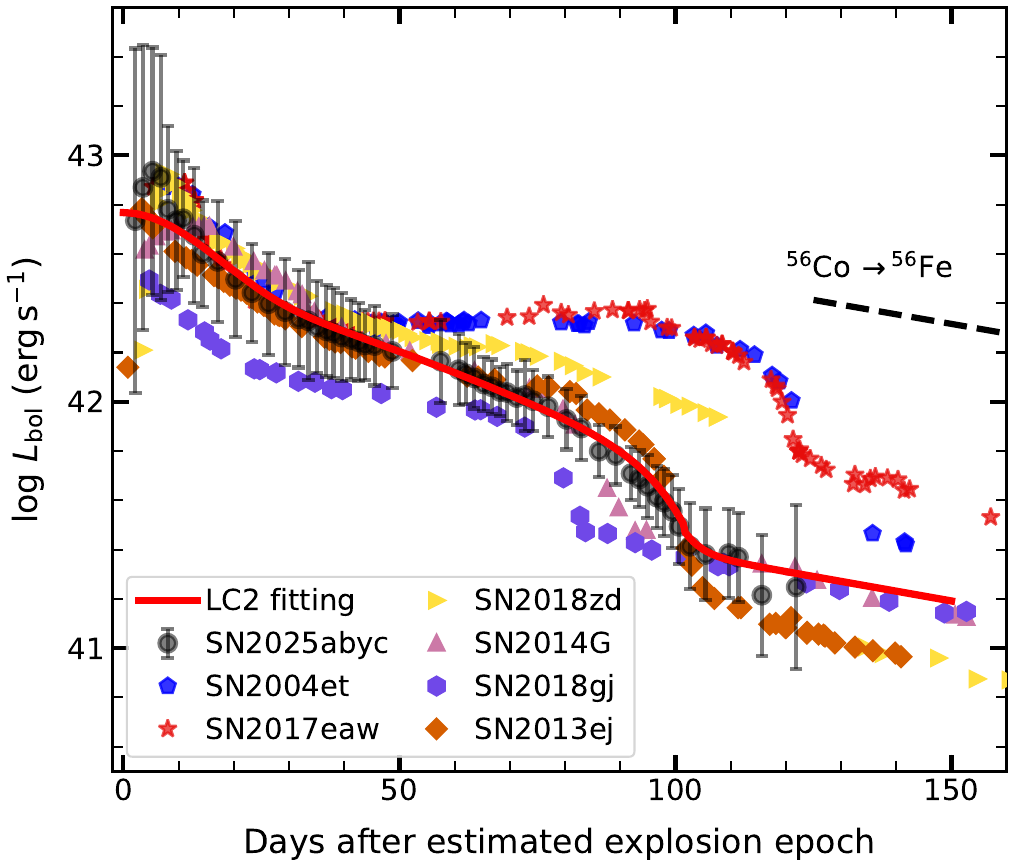}
   \caption{The evolution of bolometric luminosity of SN\,2025abyc compared with other well-studied SNe II.
   We also show two-component model fits to the SN\,2025abyc bolometric light curve, based on the model of \citet{2016A&A...589A..53N}.}
   \label{Fig:BL}
\end{figure}

\begin{figure}
  \centering
  \includegraphics[width=\linewidth]{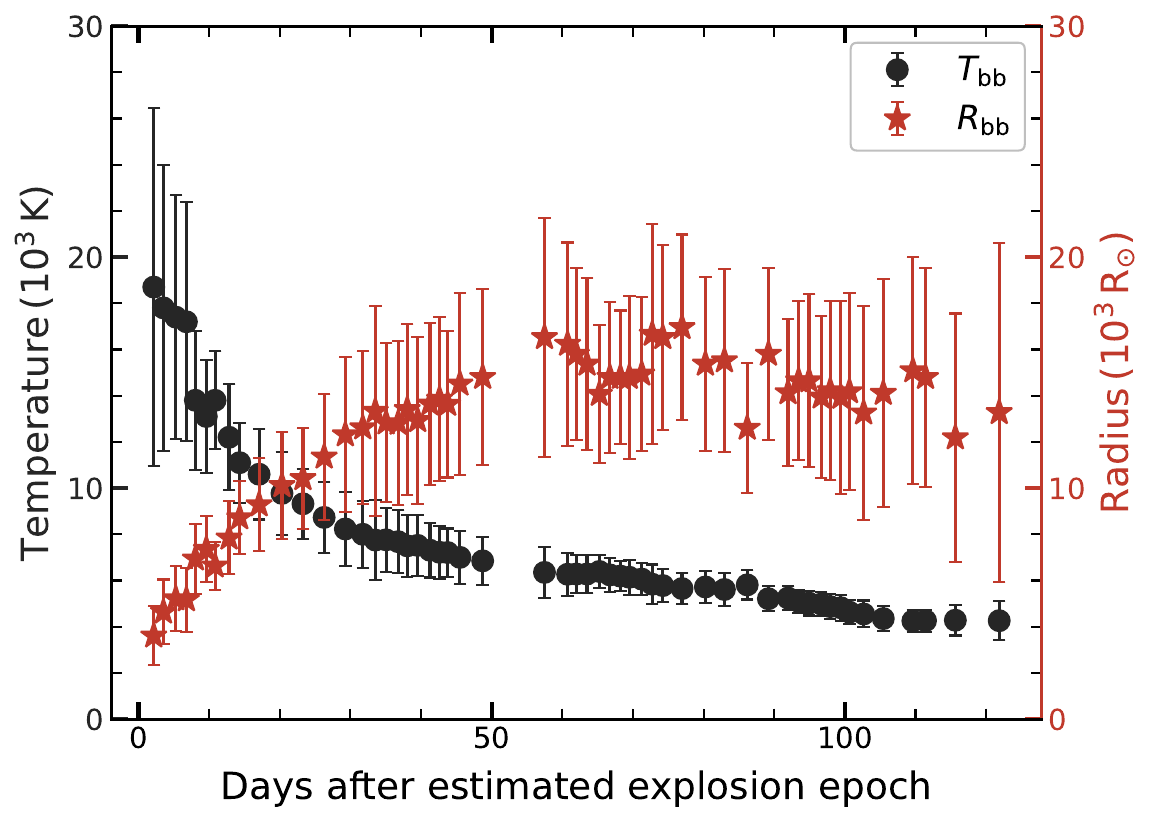}
  \caption{Evolution of the blackbody temperature $T_{\rm bb}$ and photospheric radius $R_{\rm bb}$
  inferred from the optical photometry of SN\,2025abyc based on \texttt{Superbol}.
  Phases are relative to the estimated explosion epoch.}
  \label{fig:bb_params}
\end{figure}

\begin{table}[!htbp]
    \centering
    \caption{Best-fitting parameters of the two-component model for SN\,2025abyc.} 
    \renewcommand{\arraystretch}{1.0}
     \setlength{\tabcolsep}{12pt}
    \begin{tabular*}{0.8\columnwidth}{@{\extracolsep{\fill}}ccc}
        \hline\hline
        Parameter & \multicolumn{2}{c}{SN\,2025abyc}\\
        \hline
         & Core & Shell \\
        \hline
        $R_0\ (10^{12}\,\mathrm{cm})$ & 29 & 60 \\
        $M_{\mathrm{ej}}\ (\mathrm{M}_{\odot})$ & 7.0 & 0.6 \\
        $M_{\mathrm{Ni}}\ (\mathrm{M}_{\odot})$ & 0.04 & -- \\
        $E_{\mathrm{tot}}\ (10^{51}\,\mathrm{erg})$ & 1.6 & 2.0 \\
        $E_{\mathrm{kin}} / E_{\mathrm{th}}$ & 1.67 & 10 \\
        $\kappa\ (\mathrm{cm}^2\,\mathrm{g}^{-1})$ & 0.2 & 0.4 \\
        \hline
    \end{tabular*}
    \parbox{0.9\columnwidth}{\footnotesize
    \textit{Notes:} $R_0$: initial ejecta radius; 
    $M_{\mathrm{ej}}$: ejecta mass; 
    $M_{\mathrm{Ni}}$: synthesized nickel mass; 
    $E_{\mathrm{tot}}$: total energy, including kinetic ($E_{\mathrm{kin}}$) 
    and thermal ($E_{\mathrm{th}}$) energy; 
    $\kappa$: Thomson-scattering opacity.}

    \label{tab:LC2_parameters}
    
\end{table}

\begin{figure*}
   \centering
      \sidecaption
   \includegraphics[width = 0.7\textwidth]{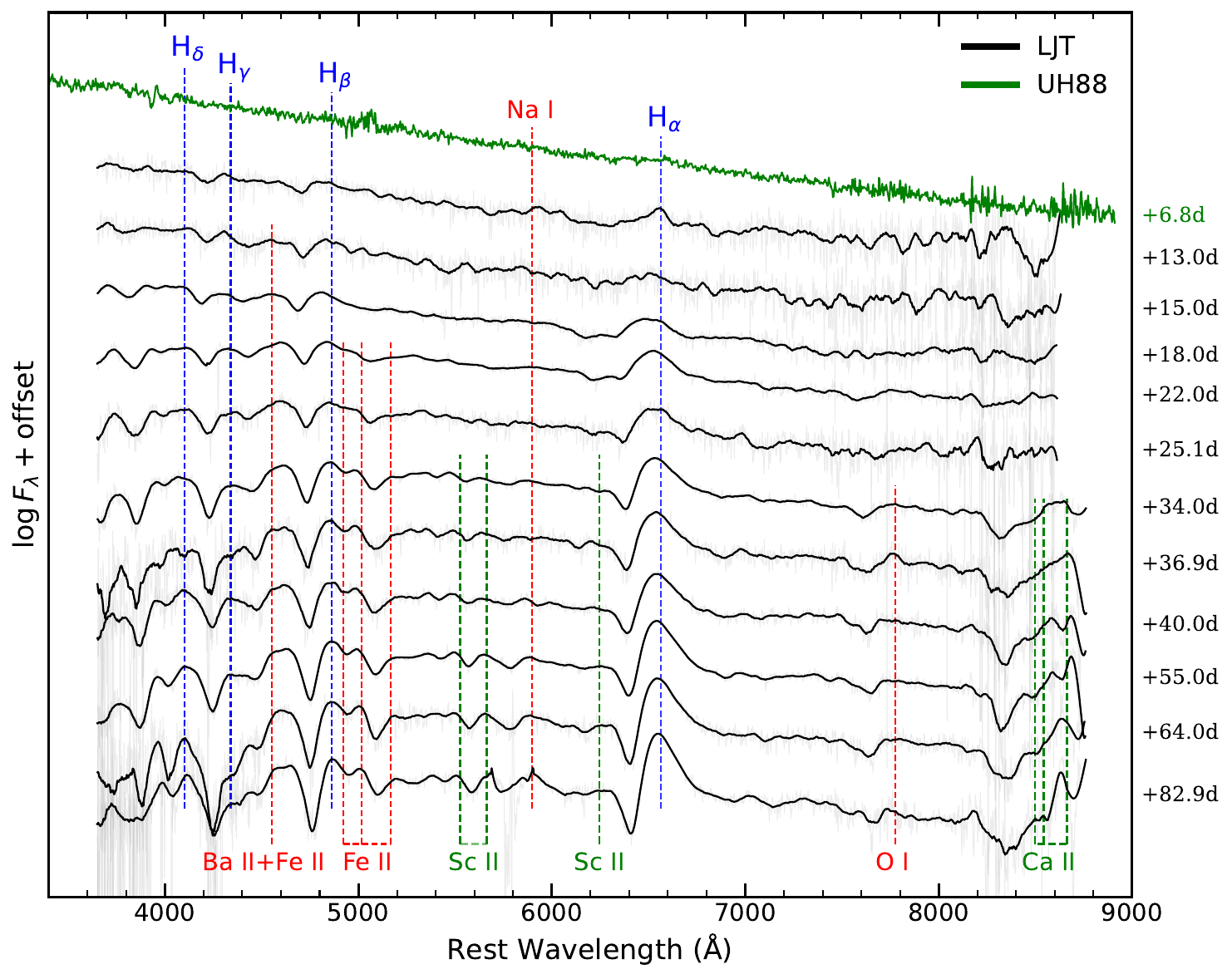}
   \caption{Spectral sequence of SN\,2025abyc.
   The phases relative to the estimated explosion epoch (MJD~60974.6) are indicated on the right-hand side. 
   All spectra obtained with LJT have been corrected for redshift and extinction. 
   Spectra with relatively low S/N ratios have been smoothed using a Savitzky–Golay filter.
   Major optical spectral lines are identified in the figure.}
    \label{Fig:spec}
\end{figure*}

For SN\,2025abyc, the bolometric light curve was constructed using the optical $gcroiz$-bands data.
The $u$ band was excluded because late-time observations are lacking and its decline is significantly faster than in the other bands, making interpolation unreliable.
We used the publicly available \texttt{Superbol} package\footnote{\url{https://github.com/mnicholl/superbol}} \citep{2018RNAAS...2..230N} to derive bolometric luminosities by integrating extinction-corrected fluxes over the observed bands.
We used the well-sampled ATLAS $o$ band as the reference band in \texttt{Superbol}.
The extrapolations beyond temporal coverage of other bands assumed a constant colour relative to the $o$ band.
After applying the adopted reddening and distance corrections, we computed the bolometric light curves of SN\,2025abyc and the comparison SNe II listed in Table~\ref{tab:SNe_II_info}, as shown in Figure~\ref{Fig:BL}.
The relevant blackbody parameters of SN\,2025abyc are shown in Figure~\ref{fig:bb_params}.
The same procedure and similar optical wavelength coverage were adopted for the comparison SNe. 
For each comparison SN, the reference band was selected as the one with the best-sampled photometric coverage.

The peak bolometric luminosity of SN\,2025abyc is higher than that of SN\,2014G and SN\,2018gj, and is comparable to those of SN\,2004et, SN\,2017eaw, and SN\,2018zd, although the uncertainty remains relatively large. 
Compared with normal SNe IIP, such as SN\,2004et and SN\,2017eaw, SN\,2025abyc exhibits a more rapid decline during the plateau phase. 
The plateau decline rate of SN\,2025abyc is more consistent with those observed in typical SNe IIL, such as SN\,2013ej and SN\,2014G. 
Furthermore, the midpoint of the plateau-to-tail transition ($t_{\rm PT}$) of SN\,2025abyc is comparable to those of these two SNe IIL. 
The radioactive tail evolution of SN\,2025abyc is similar to that of SN\,2014G and the short-plateau SN\,2018gj.

To estimate the explosion parameters of SN\,2025abyc, we fitted its bolometric light curve with the two-component model of \citet{2016A&A...589A..53N}.
This model assumes a dense inner He-rich core and an extended low-mass H-rich envelope, as commonly applied to SNe IIP.
The best-fit model is shown in Figure~\ref{Fig:BL}, and the corresponding parameters are listed in Table~\ref{tab:LC2_parameters}.
Most of the inferred parameters for SN\,2025abyc are broadly consistent with those obtained for typical SNe IIP such as SN\,2012aw and SN\,2004et in \citet{2016A&A...589A..53N}.
However, reproducing the declining plateau evolution requires a comparatively low-mass extended envelope and core ejecta mass in this two-component parametrisation \citep[][]{2021ApJ...913...55H}.
We note that this bolometric luminosity fit is intended as an exploratory parametrisation; in Section~\ref{sect:discussion} we will revisit the plateau-to-tail evolution with a more detailed multi-band modelling analysis.

Figure~\ref{fig:bb_params} shows the evolution of the blackbody temperature ($T_{\rm bb}$) and photospheric radius ($R_{\rm bb}$) inferred from our multi-band photometry.
The comparatively large $T_{\rm bb}$ uncertainties at early phases largely reflect limitations in the bolometric reconstruction: the early $u$-band detections were not included when building the bolometric light curve, reducing the wavelength leverage at epochs when the spectral energy distribution is dominated by blue flux.
Independent fits to the early spectra nevertheless suggest $T \sim 13\,000$\,K at $+6.8$\,d (consistent with the lower envelope of the $T_{\rm bb}$ uncertainties), $\sim 11\,000$\,K around $+15$\,d, and $\sim 5000$--$6000$\,K after $\sim 30$\,d.
Overall, the inferred cooling trend is qualitatively similar to that obtained with the \texttt{Superbol} analysis.
The photospheric radius rises steadily at early times and appears to plateau around $30$--$40$\,d.
At later epochs, the sparse multi-band sampling leads to large $R_{\rm bb}$ uncertainties, and we cannot robustly find a clear onset of photosphere recession associated with the end of the optical plateau.

\section{Spectroscopy}\label{sect:spectroscopy}

\subsection{Spectral sequence}
The spectral evolution of SN\,2025abyc is presented in Figure~\ref{Fig:spec}.
The first public spectrum was obtained at +6.8\,d post-explosion with the 2.2\,m UH88 telescope.
It is largely featureless and is well described by a hot blackbody continuum, with a fitted temperature of $\sim 13{\,}000$\,K.
A weak excess appears around $5000\,\AA$, where no robust line identification is expected, and is therefore likely caused by noise fluctuations.
A P-Cygni-like structure is also seen near $3950\,\AA$; given the lack of a plausible identification at this phase, it is likely an artifact related to the low S/N.

Up to +13.0\,d, the spectra are dominated by a blue continuum, with a blackbody fit yielding $T \sim 11{\,}000$\,K. 
\ha\ appears weak and broad, while \hb\ and \hg\ show clearly detected P-Cygni structure.
The spectra at +13.0\,d and +15.0\,d have relatively low S/N and do not exhibit clear \ha\ P-Cygni profiles.
This morphology may be related to the normal early-time spectroscopic evolution of SNe~II, during which the \ha\ profile is often dominated by emission \citep{2017ApJ...850...89G,2019MNRAS.490.2799D}.
Alternatively, ongoing CSM interaction may have contributed to filling in or masking the absorption trough \citep{2018MNRAS.476.1497B,2023A&A...677A.105D}.
By +18.0\,d, a well-developed P-Cygni profile emerges in \ha.
At later epochs, Balmer lines consistently show prominent P-Cygni morphology, and their absorption-line velocities decline gradually with time.

As the ejecta expand and cool through the plateau phase, the continuum temperature decreases.
Numerous metal lines, including \Feii, \Scii, \Baii, and \Caii, progressively emerge and strengthen.
A possible \Oi $\lambda7774$ feature is also present; however, this region is strongly affected by telluric absorption, and residual contamination may remain despite the telluric correction. 
These features become narrower and deeper with time, and their absorption minima move toward rest wavelengths, consistent with the photosphere receding into slower-moving ejecta.

The photometric light curve in the $croi$ bands shows an evolving decline rate, with the most pronounced change occurring $\sim$70 days after explosion.
Between approximately 30 and 70\,d, the spectra show only modest changes, mainly manifested as a slight strengthening of the metal lines, while the Balmer-line profiles and absorption velocities evolve relatively smoothly.
By $+82.9$\,d, the spectrum shows no abrupt change in the measured line velocities, which continue to decline monotonically (Figure~\ref{Fig:line_velocity}).
Thus, no abrupt spectroscopic change is observed around the photometrically identified onset of the plateau-to-tail transition, although the available spectral sampling is limited.

\subsection{Comparison with other SNe II}
\begin{figure*}
   \centering   \sidecaption
   \includegraphics[width = 0.7 \linewidth]{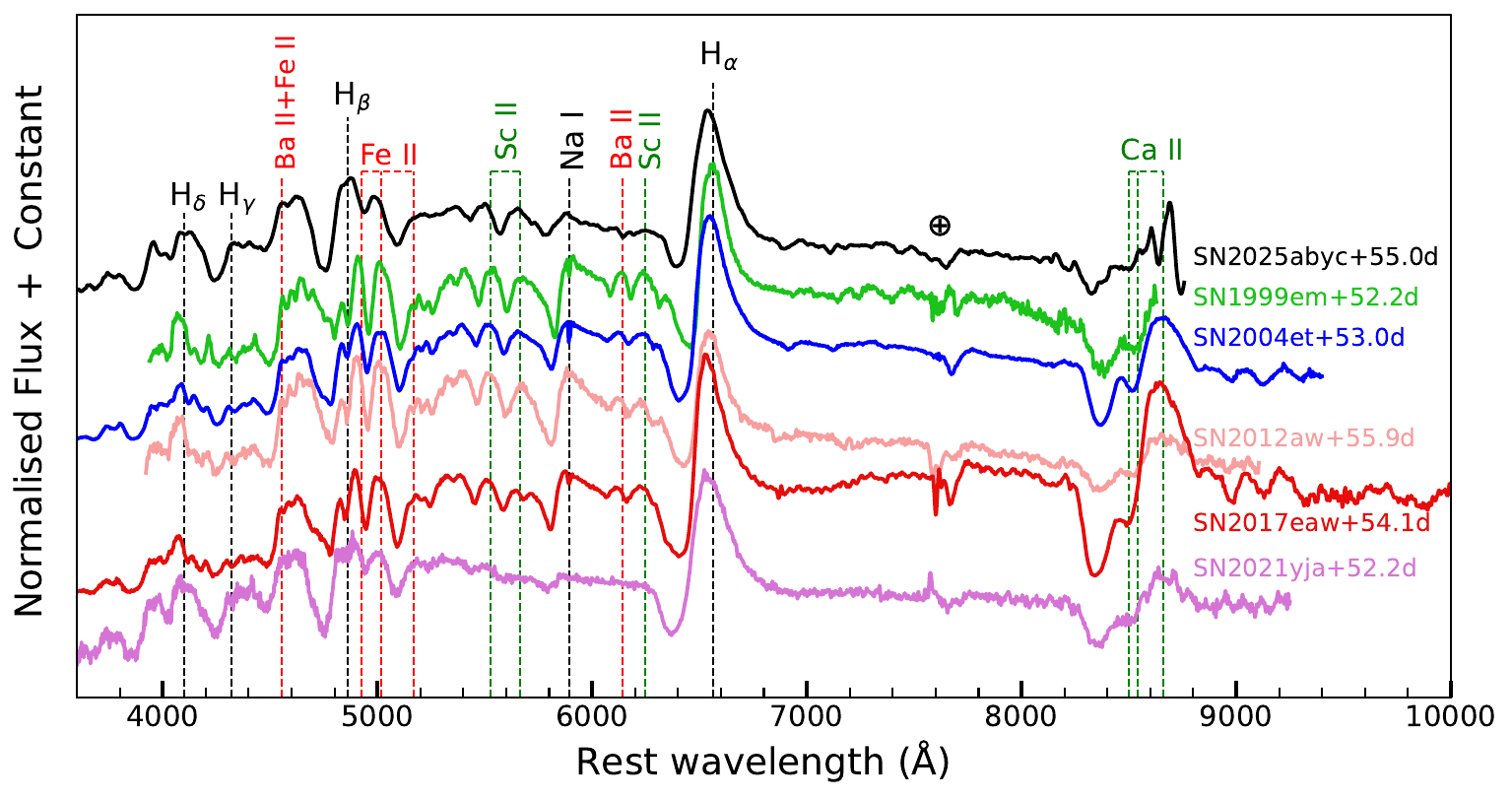}
   \caption{Intermediate-phase spectroscopic comparisons of SNe IIP.
    Phases relative to the explosion epoch are indicated on the right-hand side, and all spectra have been corrected to the rest frame.
    Major spectral features are identified and labeled.
    }
    \label{Fig:spec_com}
\end{figure*}

\begin{figure}
   \centering
   \includegraphics[width = 1 \linewidth]{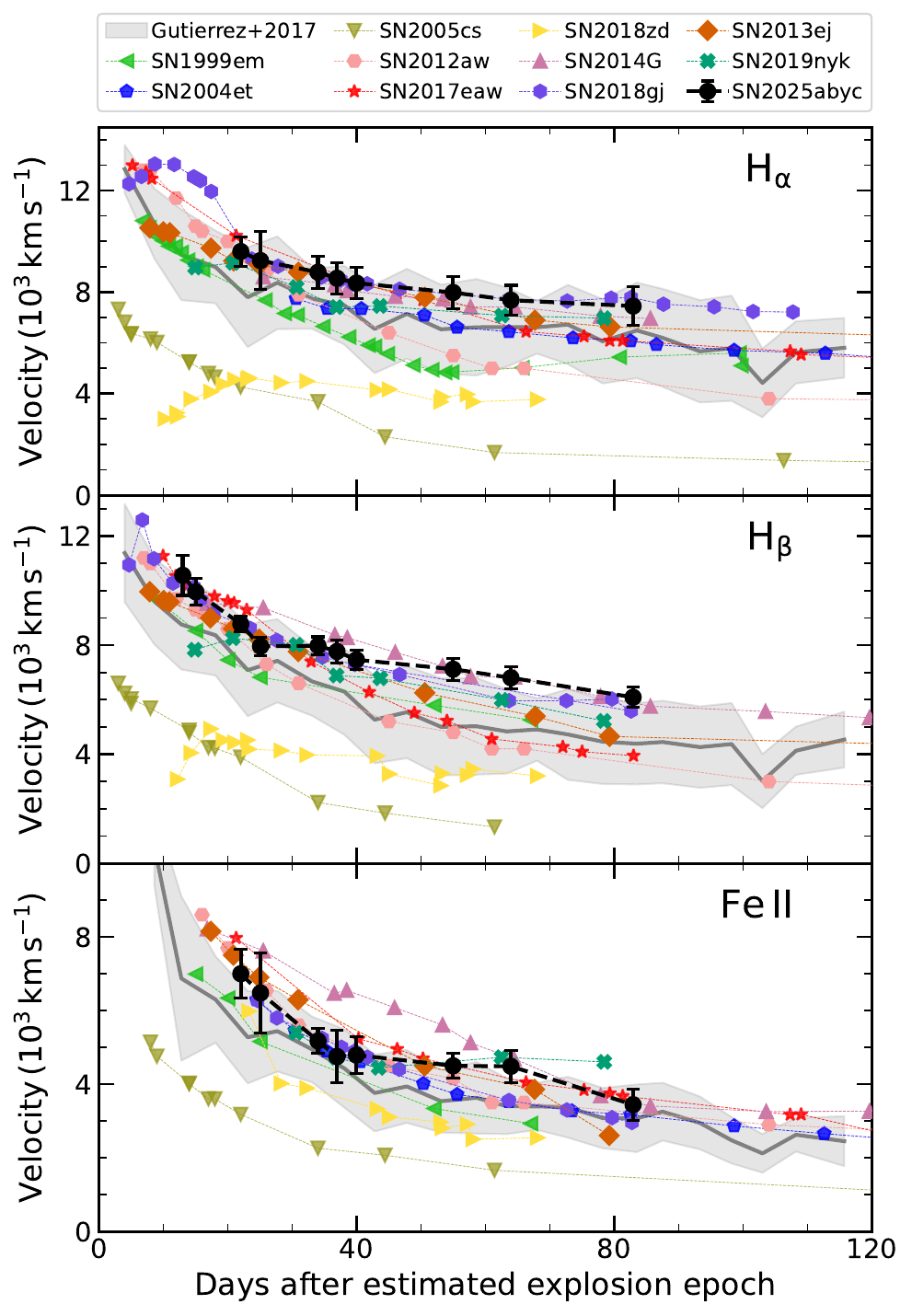}
   \caption{Evolution of the line velocities of \ha, \hb, and \Feii~$\lambda\,5169$ for SN\,2025abyc compared with those of several well-studied SNe II. 
   The velocities are measured from the minima of the P-Cygni absorption components. 
   The gray shaded region represents the average velocity evolution of a sample of SNe II from \citet{2017ApJ...850...89G}.}
    \label{Fig:line_velocity}
\end{figure}

Figure~\ref{Fig:spec_com} compares the spectrum of SN\,2025abyc at $\sim$50\,d post-explosion with those of other SNe IIP at similar phases.
At this epoch, all objects are dominated by strong \ha\ P-Cygni profiles, although the line width and absorption depth vary from event to event.
For SN\,2025abyc, the \ha\ absorption velocity is comparable to those of SN\,2004et and SN\,2017eaw.

In addition to Balmer lines, multiple metal features are visible, including \Feii, \Baii, \Scii, and \Caii, especially at shorter wavelengths.
The \Feii\ and \Scii\ absorptions in SN\,2025abyc appear relatively blueshifted compared with several comparison objects at similar phases, suggesting comparatively high line velocities at this stage.
Overall, the plateau-phase spectral morphology of SN\,2025abyc is most similar to that of SN\,2004et.
The near-infrared \Caii\ triplet appears narrower and weaker than in our comparison objects, which reflects a lower optical depth and/or a smaller velocity range contributing to the line formation.
The \Oi\ $\lambda7774$ feature is difficult to identify reliably in SN\,2025abyc because it lies in a spectral region strongly affected by telluric absorption.
Although the spectra were corrected for telluric absorption, residual contamination may remain.
For comparison, this feature is more clearly detected in SN\,2004et and SN\,2017eaw.

\subsection{Spectral line velocity evolution}

The velocity evolution of \ha, \hb, and \Feii~$\lambda5169$ for SN\,2025abyc is shown in Figure~\ref{Fig:line_velocity}.
Line velocities were measured from the absorption minima of the P-Cygni profiles in rest-frame spectra.
For each feature and epoch, we fitted the local profile with a two-component Gaussian model using MCMC and adopted the posterior median as the velocity.
For reference, we overplot the mean trend and dispersion of the SN II sample from \citet{2017ApJ...850...89G}, together with several well-studied SNe II (as shown in Figure~\ref{Fig:line_velocity}).

At early epochs, the Balmer-line velocities of SN\,2025abyc are comparable to those of typical SNe IIP, such as SN\,2012aw and SN\,2017eaw, and are close to or slightly higher than those of SNe IIL, including SN\,2013ej, SN\,2014G, and SN\,2019nyk.
At $+13$\,d, the \hb\ absorption velocity reaches $10\,600$\,km\,s$^{-1}$, placing it near the upper end of the mean distribution, $8760 \pm 1650$\,km\,s$^{-1}$ \citep{2017ApJ...850...89G}.
By $+22$\,d, the \ha\ and \hb\ velocities are $\rm 9600\,km\,s^{-1}$ and $8800$\,km\,s$^{-1}$, respectively, again exceeding the corresponding sample means of $9000\pm1400\rm \,km\,s^{-1}$ and $\rm 7100 \pm 1700$\,km\,s$^{-1}$. 
At $+55$\,d, the \Feii\ velocity is $4500$\,km\,s$^{-1}$, compared with $3500 \pm 850$\,km\,s$^{-1}$ for the sample. 
During intermediate epochs, the decline of the Balmer-line velocities appears relatively shallow, keeping SN\,2025abyc close to the upper envelope of the comparison distribution.
The overall ejecta velocities of SN\,2025abyc are comparable to those of SN\,2013ej, SN\,2014G, and SN\,2018gj, all of which are located toward the higher-velocity end of the SN II population.
We note that the uncertainty in host-galaxy redshift introduces an additional systematic uncertainty in the absolute velocity scale.
However, this systematic offset is common to all epochs and therefore does not affect the relative velocity evolution.
Thus, SN\,2025abyc follows the generic velocity-decline trend of SNe II, while generally remaining toward the high-velocity side of the comparison sample throughout the observed photospheric phase.

\section{Discussion} \label{sect:discussion}

\subsection{Comparison with ZTF samples}
\begin{figure}
    \centering
    \includegraphics[width=0.98\linewidth]{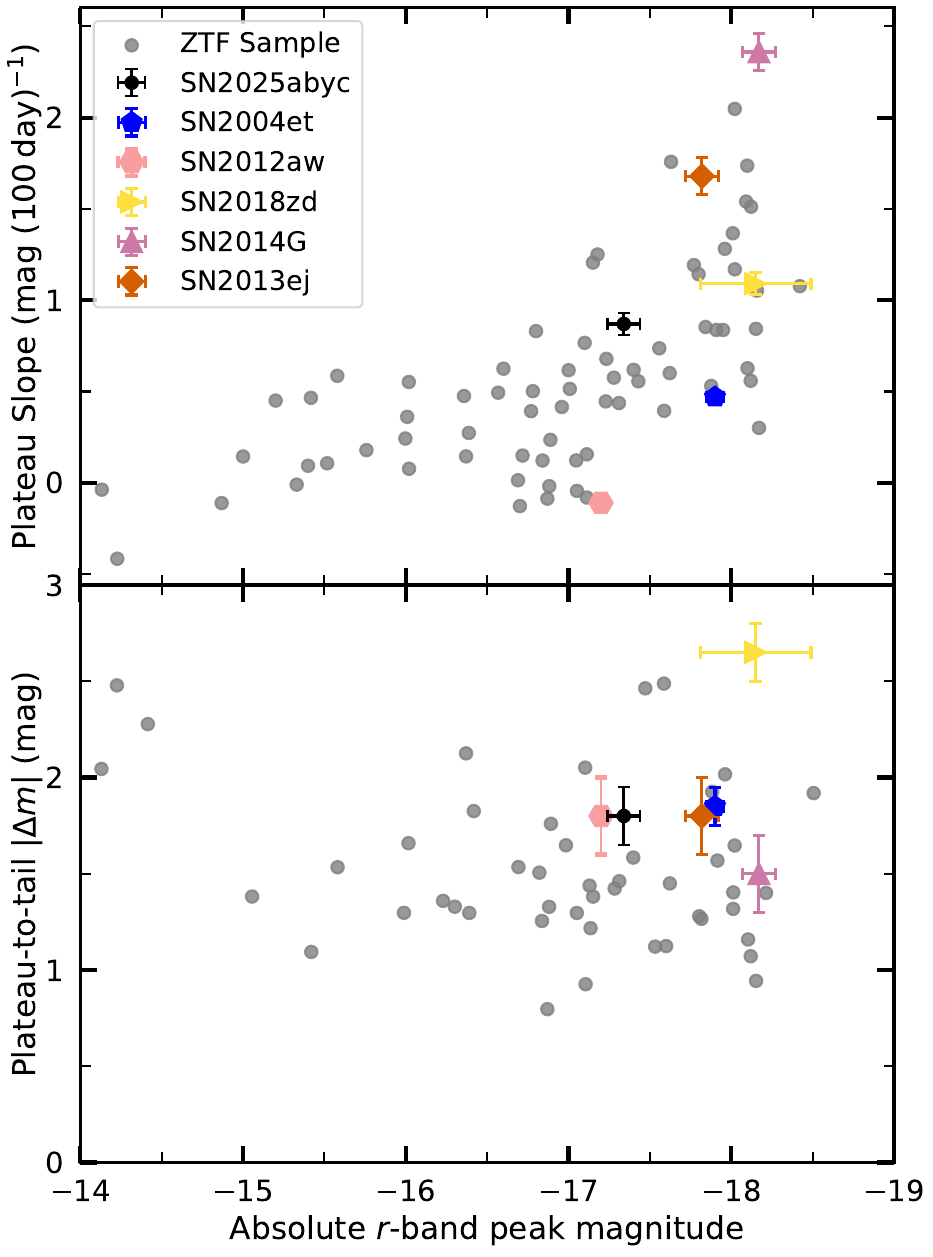}
    \caption{Comparison of the plateau slope and plateau$-$tail magnitude drop as a function of the peak absolute magnitude in the $r$ band. 
    The ZTF samples are adopted from \citet{2026PASP..138b4204D}. }
    \label{fig:comparisons}
\end{figure} 

\begin{figure*}[t]
    \centering
    \begin{minipage}{0.49\textwidth}
        \centering
        \includegraphics[width=\linewidth]{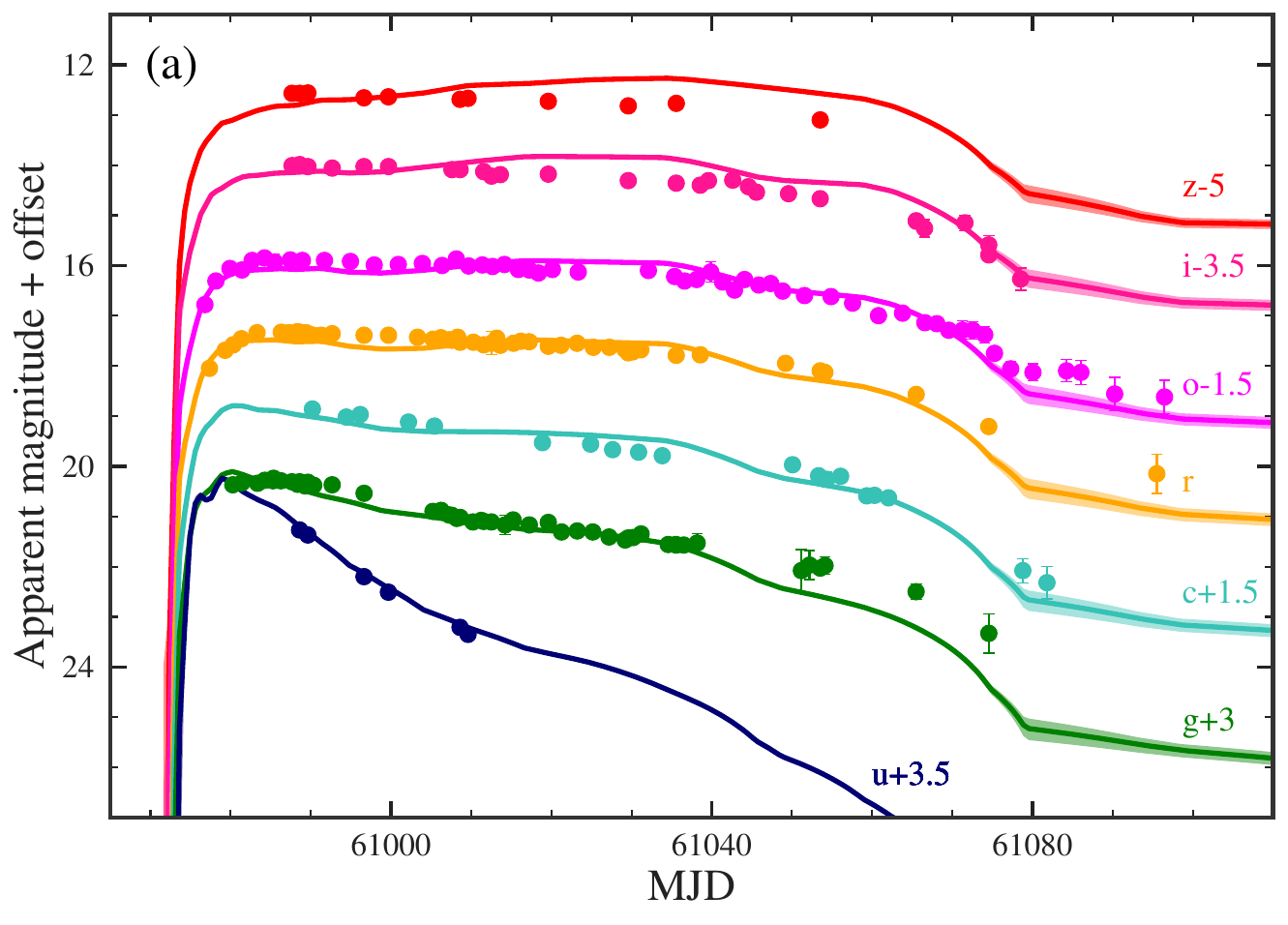}
    \end{minipage}
    \hfill
    \begin{minipage}{0.49\textwidth}
        \centering
        \includegraphics[width=\linewidth]{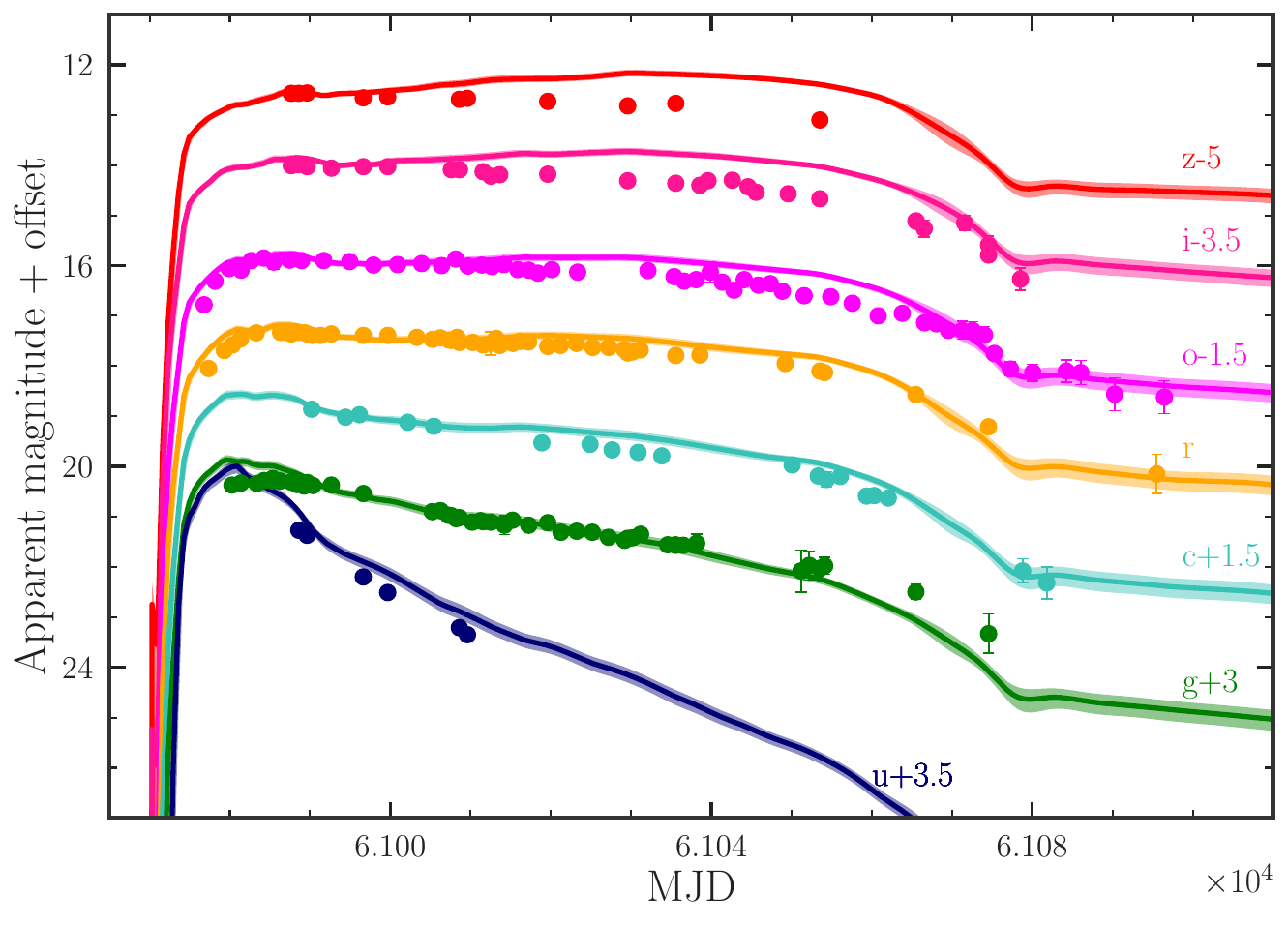}
    \end{minipage}
    \caption{\textit{Left panel:} multi-band light-curve fit of SN\,2025abyc with the surrogate model.
            \textit{Right panel:} multi-band fit using the GPU-accelerated machine-learning reconstruction of the surrogate model.
            The shaded region indicates the maximum-likelihood realization, including a $3\sigma$ systematic uncertainty.}
    \label{fig:multi-band_fitting}
\end{figure*}

To place SN\,2025abyc in a broader population context, we compare its plateau decline rate with the peak absolute $r$-band magnitude, $M_r$, in the $M_r$--slope plane (upper panel of Figure~\ref{fig:comparisons}), together with the ZTF SN~IIP sample from \citet{2026PASP..138b4204D}.
In the upper panel, SN\,2025abyc lies close to the main distribution of the sample, exhibiting a comparable $M_r$ and a plateau slope slightly above the sample average. 
Nevertheless, its decline rate remains lower than those of the representative SNe IIL, including SN\,2013ej and SN\,2014G.

The lower panel of Figure~\ref{fig:comparisons} shows the plateau-to-tail magnitude drop as a function of $M_r$.
We note that the plateau-to-tail magnitude drop ($\Delta m$) is plotted as $m_{\rm plateau} - m_{\rm tail}$ (negative for a fading SN).
Here we quote $|\Delta m|$ for readability.
For the ZTF sample, $\Delta m$ is measured in the $r$ band and only SNe with a clearly detected tail are included \citep{2026PASP..138b4204D}.
For SN\,2025abyc, we adopt the ATLAS $o$ band because of its superior late-time coverage and measure a magnitude difference of $\Delta m\approx1.8$\,mag between 70 and 110\,d, as shown in Figure~\ref{fig:phase}.
As shown in Figure~\ref{fig:Abs_r}, the magnitude drops measured in the $o$ and $r$ bands are consistent within $\lesssim 0.1$\,mag.
We note that this directly measured magnitude difference is not equivalent to the Fermi--Dirac parameter $a_0$, which represents the asymptotic transition amplitude after accounting for the underlying linear evolution around $t_{\rm PT}$ ($\sim 100.5\,$d for SN\,2025abyc).
Most SNe~II in the comparison sample show plateau-to-tail drops of $\sim 1$--$2.5$\,mag.
Among these events, the ECSN candidate SN\,2018zd lies toward the upper end of this range.
The magnitude drop of SN\,2025abyc is slightly larger than the sample average, although it remains within the scatter of the population. 
Compared with the ZTF sample median of $1.3^{+0.4}_{-0.5}$\,mag reported by \citet{2026PASP..138b4204D}, SN\,2025abyc shows a relatively larger drop, while its value is comparable to those measured for SN\,2012aw, SN\,2013ej, and SN\,2004et.

\subsection{Surrogate model}

\citet{2023PASJ...75..634M} presented a grid of SN surrogate models from RSG explosions computed with the one-dimensional radiation-hydrodynamics code \texttt{STELLA}.
The grid spans a range of input parameters, including progenitor masses, nickel mass, mass-loss rates, and CSM properties such as radii and density structure.
Building on this work, \citet{2025MNRAS.544.2653S} developed surrogate models that enable rapid inference against observations.
They made these surrogate models and an associated user interface publicly available through the open-source \texttt{redback} package \citep{2024MNRAS.531.1203S}.
\citet{2025MNRAS.544.2653S} also demonstrated multi-band light-curve fits for typical SNe IIP, including SN\,2004et, SN\,2012aw, and SN\,2017gmr.

Subsequently, we attempted to fit SN\,2025abyc using the surrogate model implemented in \texttt{redback} 
\footnote{See redback.transient\_models.supernova\_models.typeII\_bolometric at \url{https://redback.readthedocs.io/en/latest}}.
We adopted prior ranges consistent with those of \citet{2025MNRAS.544.2653S}, except for the nickel mass, for which we used a narrower interval of $0.027$--$0.050\,M_\odot$ informed by the approximate $M_{\mathrm{Ni}}\sim 0.04\,M_\odot$ suggested by our earlier two-component bolometric fit in Section~\ref{sect:Bol}.
The multi-band fitting results are shown in the left panel of Figure~\ref{fig:multi-band_fitting}, and the posterior distributions are summarized in the left panel of Figure~\ref{fig:Corner}.

Overall, the surrogate model broadly reproduces the gradual steepening in the light-curve evolution and the prolonged plateau-to-tail transition, particularly in the $cro$ bands.
The model captures an increase in the decline rate around MJD$\sim$61035, corresponding to $\sim$60\,d after explosion in Figure~\ref{Fig:LC}.
We also note that the standard-model $g$-band light curve shown by the black line in Figure~3 (panel~1) of \citet{2023PASJ...75..634M} exhibits a qualitatively similar change in the plateau decline rate at phases of $\sim$50\,d after explosion.
Such behaviour could in principle reflect a combination of zero-age main-sequence (ZAMS) mass ($M_{\rm ZAMS}$), explosion energy ($E_{\rm sn}$), and mass-loss rate ($\dot{M}$).
Additionally, the fit is not equally good at all epochs and bands: at late times the predicted $g$-band light curve lies systematically below the observations, while the $z$-band prediction is too bright.
In addition, the posterior distributions summarized in Figure~\ref{fig:Corner} show that the inferred ZAMS mass is pushed toward the lower edge of the adopted prior ($10\,M_\odot$), while $R_{\mathrm{CSM}}$ is driven toward the upper prior bound ($\rm  10^{15}\,cm$).
Therefore, although the surrogate model provides a reasonable qualitative match to the observed plateau morphology, the inferred physical parameters should be interpreted with caution given the likely model-dependent degeneracies and boundary effects from the priors.

Subsequently, we adopted an alternative surrogate model with a more extended parameter grid than that of \citet{2023PASJ...75..634M}. 
This model combines an autoencoder with an emulator, enabling a more accurate reconstruction of the light curves \citep{2026PhRvD.114b3020Z}.
The resulting fits and parameter constraints are presented in the right panel of Figure~\ref{fig:multi-band_fitting} and Figure~\ref{fig:Corner}.
Relative to the baseline surrogate model, this reconstruction provides a noticeably better match to the late-time tail evolution, particularly in the $c$, $r$, and $o$ bands.
The inferred nickel mass is $M_{\mathrm{Ni}}\sim0.033\,M_\odot$, marginally higher than the value obtained from the standard surrogate fit ($M_{\mathrm{Ni}}\sim 0.030\,M_\odot$).
The inferred ZAMS progenitor mass is $11.1\,M_\odot$, which is no longer pinned to the lower edge of the adopted prior, suggesting reduced sensitivity to prior boundary effects in this framework.
Nevertheless, both frameworks yield a relatively large value of $R_{\mathrm{CSM}}$, which mainly controls the early rise ($\lesssim 20$\,d before peak), where larger CSM radii tend to produce a higher and earlier peak luminosity.

\subsection{Possible explanations}

The gradual departure from the approximately linear plateau evolution at $\sim70$\,d and the subsequent prolonged evolution toward the radioactive tail may reflect several physical effects.
We discuss three possible explanations below, which are not necessarily mutually exclusive.

(1) Interaction with an extended or structured CSM may modify the light-curve evolution.
During the early phases, CSM interaction can provide additional luminosity, while a gradual decrease in the interaction contribution as the ejecta propagate through a declining CSM density profile could alter the subsequent decline rate.
In the surrogate model, the parameter $\beta$ regulates the steepness of the CSM density profile and, together with the mass-loss rate and CSM radius, affects the resulting light-curve morphology \citep{2023PASJ...75..634M}.
Both surrogate-model frameworks favour a relatively extended CSM, although the inferred CSM parameters are affected by model degeneracies and prior-boundary effects.
The weak and broad \ha\ profile observed at $+13$\,d is consistent with partial filling of the absorption trough by CSM-interaction emission.
However, this morphology can also arise from the normal early spectroscopic evolution of SNe II, and therefore does not provide unambiguous evidence for CSM interaction.

(2) The distribution of $^{56}$Ni may influence the timing and morphology of the plateau-to-tail transition.
\citet{2013MNRAS.434.3445P} explored different degrees of central nickel mixing using the parameter $K_{\mathrm{mix}}$, where larger $K_{\mathrm{mix}}$ corresponds to a more centrally concentrated $^{56}$Ni distribution.
In such a configuration, radioactive heating contributes less efficiently to the outer ejecta during the plateau and becomes important at later phases.
This delayed contribution can modify the curvature of the late plateau and the subsequent evolution toward the radioactive tail (see their Figure~11, model~25\_k1).

(3) The gradual transition may be related to the density structure of the hydrogen-rich envelope.
The plateau slope is sensitive to the envelope density profile: a flatter and more extended structure tends to maintain a nearly constant photospheric radius, whereas a steeper profile produces a more rapidly declining light curve \citep{2009ApJ...703.2205K,2019ApJ...879....3G}.
As the recombination front recedes through the expanding envelope, a gradual change in the density gradient could alter its recession rate and produce a progressive steepening of the light curve \citep{2011ApJ...729...61B,2013MNRAS.433.1745D}.
In this interpretation, the departure from the approximately linear plateau at $\sim70$\,d marks the onset of the gradual plateau-to-tail evolution, while $t_{\rm PT}\approx100.5$\,d represents the midpoint of this transition.

The current photometric and spectroscopic data do not uniquely distinguish among these scenarios.
The prolonged plateau-to-tail evolution of SN\,2025abyc may therefore result from a combination of CSM properties, $^{56}$Ni distribution, and hydrogen-envelope structure.

\section{Conclusion} \label{sect:conclusion}

The photometric evolution of SN\,2025abyc does not fit cleanly within the traditional IIP--IIL classification.
Its relatively rapid plateau decline at blue wavelengths is IIL-like, whereas its red-band light curves evolve more slowly and occupy an intermediate position between the canonical IIP and IIL comparison samples.
The most distinctive feature of SN\,2025abyc is the gradual manner in which it approaches the radioactive tail.
The light curve begins to depart from its nearly linear plateau evolution approximately 30\,d before the fitted transition midpoint based on the Fermi--Dirac function, producing a morphologically prolonged transition relative to the comparison events.
SN\,2025abyc otherwise follows the general spectroscopic evolution of SNe II, although its expansion velocities remain toward the high-velocity side of the observed distribution.
Combining the two-component bolometric fit with the surrogate light-curve modelling, we infer a synthesised nickel mass in the range $M_{\mathrm{Ni}}\sim 0.03$--$0.04\,M_\odot$.
The surrogate-model fits broadly reproduce the gradual evolution from the plateau toward the radioactive tail.
The alternative surrogate-model reconstruction further suggests a ZAMS progenitor mass of $M_{\mathrm{ZAMS}}\sim 11\,M_\odot$, which is no longer pinned to the prior boundaries in our inference.
Within these model frameworks, the fits favour a relatively extended CSM environment, with a large $R_{\mathrm{CSM}}$ required to reproduce the early rise of the light curve.
We suggest that the gradual plateau-to-tail evolution may arise from the circumstellar environment, the distribution of $^{56}$Ni, or the density structure of the hydrogen-rich envelope.

\section*{Data availability}
The majority of photometric data are presented in tables in the Appendix.
Our spectral observations are available via the Weizmann Interactive Supernova Data Repository \citep[WISeREP;][]{2012PASP..124..668Y} at \url{https://www.wiserep.org/object/29325}.

\begin{acknowledgements}
We gratefully thank the anonymous referee for his/her insightful comments and suggestions that improved the paper. 

This study is supported by the CAS Project for Young Scientists in Basic Research (YSBR-148), the National Natural Science Foundation of China (Nos 12225304, 12288102), the National Key R\&D Program of China (No. 2021YFA1600404), the Yunnan Revitalization Talent Support Program (Yunling Scholar Project), the Yunnan Science and Technology Program (Nos 202501AS070005, 202605AS350010, and 202601BC070011), and the International Centre of Supernovae (ICESUN), Yunnan Key Laboratory of Supernova Research (No. 202505AV340004). 
J.Z. is supported by the B-type Strategic Priority Program of the Chinese Academy of Sciences (Grant No. XDB1160202), the National Natural Science Foundation of China (NSFC grants 12173082 and 12333008),  the Yunnan Fundamental Research Projects (YFRP; grants 202501AV070012 and 202401BC070007).

We acknowledge the support of the staff of the LJT and the MATCH.
Funding for the LJT has been provided by the CAS and the People’s Government of Yunnan Province. 
The LJT is jointly operated and administrated by YNAO and Center for Astronomical Mega-Science, CAS.

Based on observations obtained with the Samuel Oschin Telescope 48-inch and the 60-inch Telescope at the Palomar Observatory as part of the Zwicky Transient Facility project.
ZTF is supported by the National Science Foundation under Grants No. AST-1440341 and AST-2034437 and a collaboration including current partners Caltech, IPAC, the Oskar Klein Center at Stockholm University, the University of Maryland, University of California, Berkeley, the University of Wisconsin at Milwaukee, University of Warwick, Ruhr University, Cornell University, Northwestern University and Drexel University. Operations are conducted by COO, IPAC, and UW.

This work has made use of data from the Asteroid Terrestrial-impact Last Alert System (ATLAS) project. The Asteroid Terrestrial-impact Last Alert System (ATLAS) project is primarily funded to search for near earth asteroids through NASA grants NN12AR55G, 80NSSC18K0284, and 80NSSC18K1575; byproducts of the NEO search include images and catalogs from the survey area. This work was partially funded by Kepler/K2 grant J1944/80NSSC19K0112 and HST GO-15889, and STFC grants ST/T000198/1 and ST/S006109/1. The ATLAS science products have been made possible through the contributions of the University of Hawaii Institute for Astronomy, the Queen’s University Belfast, the Space Telescope Science Institute, the South African Astronomical Observatory, and the Millennium Institute of Astrophysics (MAS), Chile.

\end{acknowledgements}

\bibliographystyle{aa}
\bibliography{aa61397-26}

\begin{appendix}
\section{Supplementary figures}

\begin{figure}[t]
   \centering
   \includegraphics[width = \columnwidth]{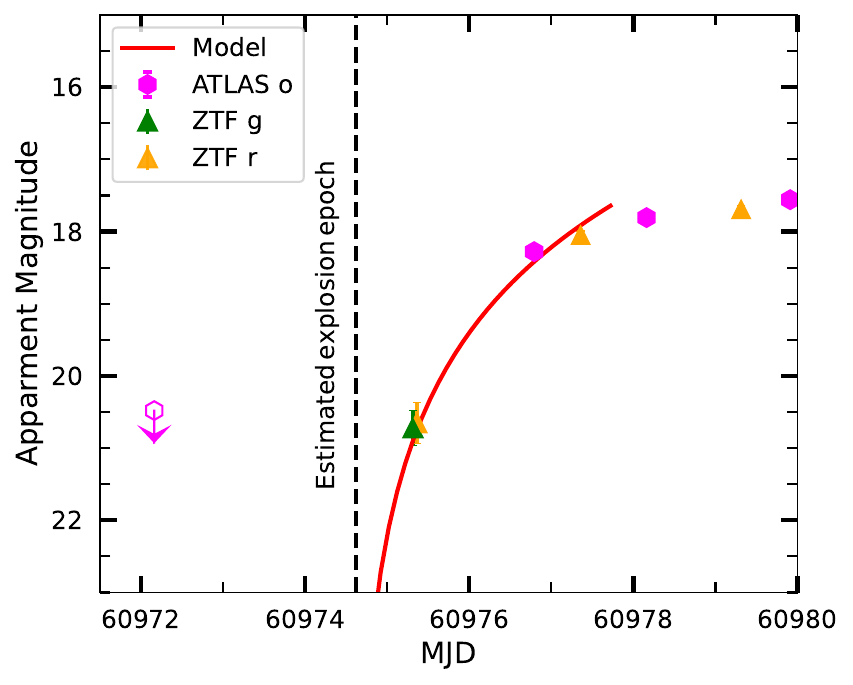}
      \caption{Expanding fireball model fit to the early detections of SN\,2025abyc in the ATLAS $o$ band and ZTF $gr$ band. 
      The ATLAS forced-photometry exposures were combined into nightly stacks.
      The two earliest significant ZTF detections, at  MJD~60975.3, were obtained from forced photometry, whereas the subsequent ZTF points are standard alert-stream detections.}
     \label{Fig:explosion_epoch}
\end{figure}

\begin{figure}[t]
   \centering
   \includegraphics[width = \columnwidth]{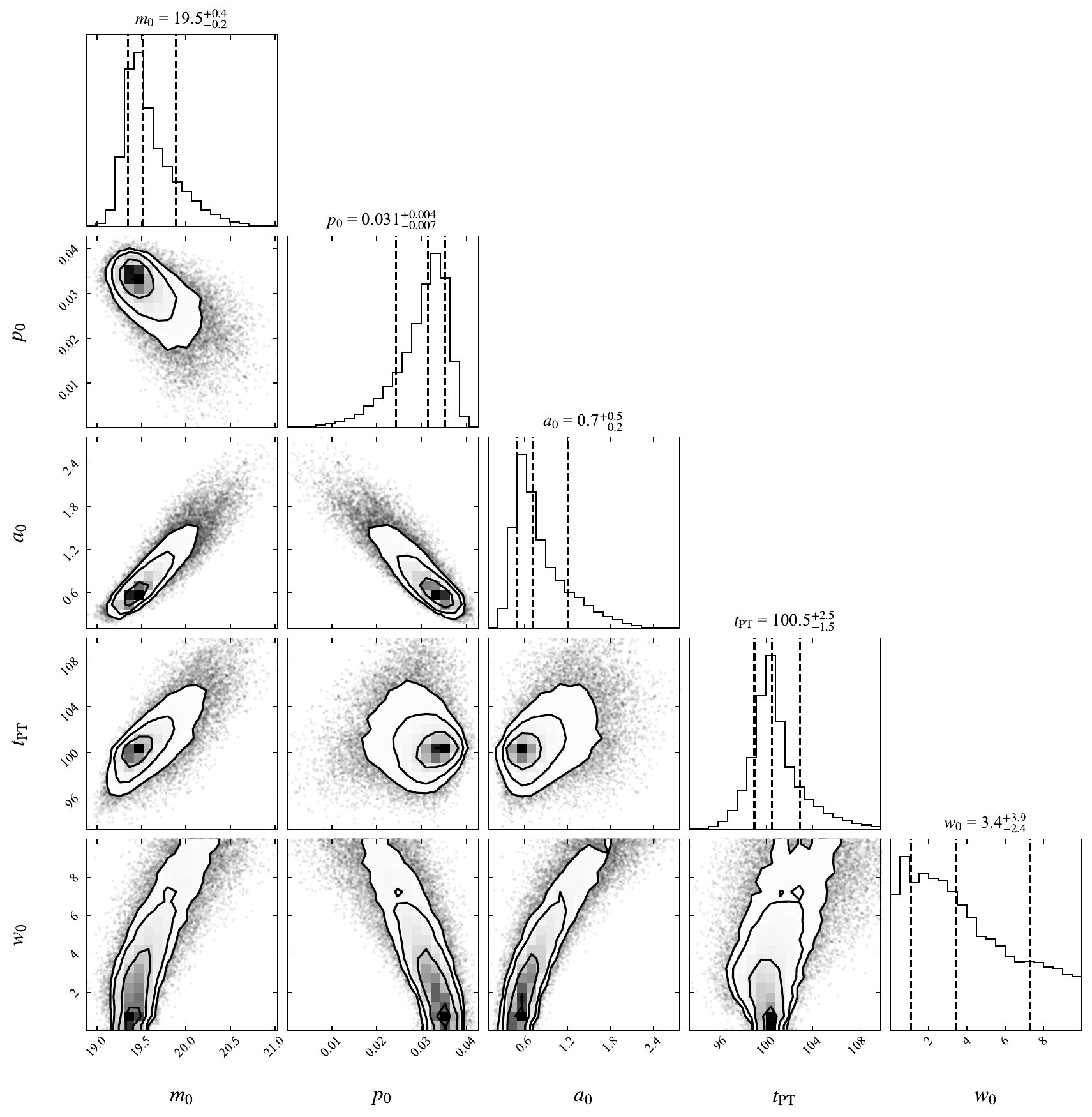}
      \caption{
      Posterior corner plot for SN\,2025abyc $o$-band data during the transition phase (after $\sim$70\,d) based on the Fermi--Dirac function.}
     \label{Fig:FDcorner}
\end{figure}

\begin{figure*}[t]
   \centering
   \includegraphics[width = \textwidth]{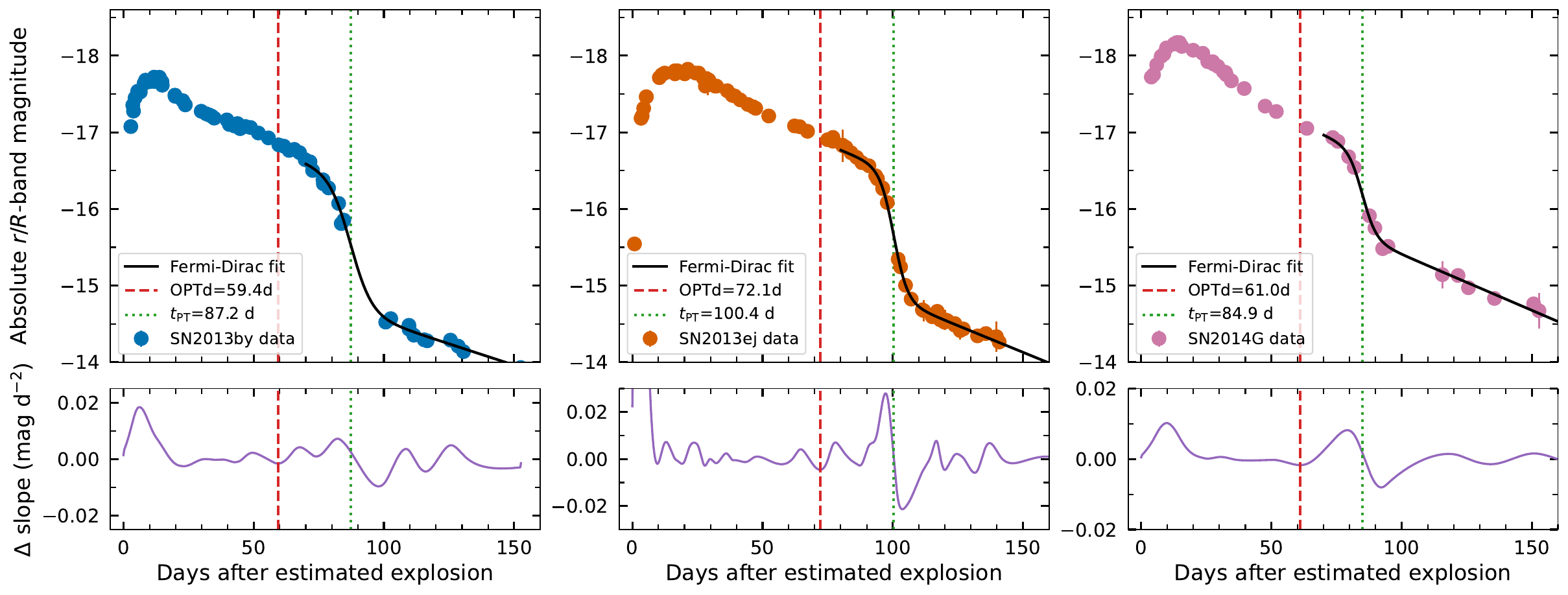}
      \caption{Absolute $r/R$-band light curves of SN\,2013by, SN\,2013ej, and SN\,2014G, and their Fermi--Dirac function fits.
      The lower panels show the evolution of the $\Delta$ slope calculated from the light curves after applying Gaussian Process regression \citep{7130620}.
      The $t_{\rm PT}$ values are directly determined from the Fermi–Dirac fits and are marked by the green dotted lines. 
      The OPTd values are identified between 40 days after explosion and $t_{\rm PT}$, when the light curves begin to decline steeply. 
      Following the method of \citet{2026PASP..138b4204D}, OPTd is defined as the epoch at which $\Delta$ slope reaches its minimum within this time interval, corresponding to the point of the most rapid transition.
}
     \label{Fig:SNe_drop}
\end{figure*}

\begin{figure*}[t]
    \centering
    \includegraphics[width=0.48\linewidth]{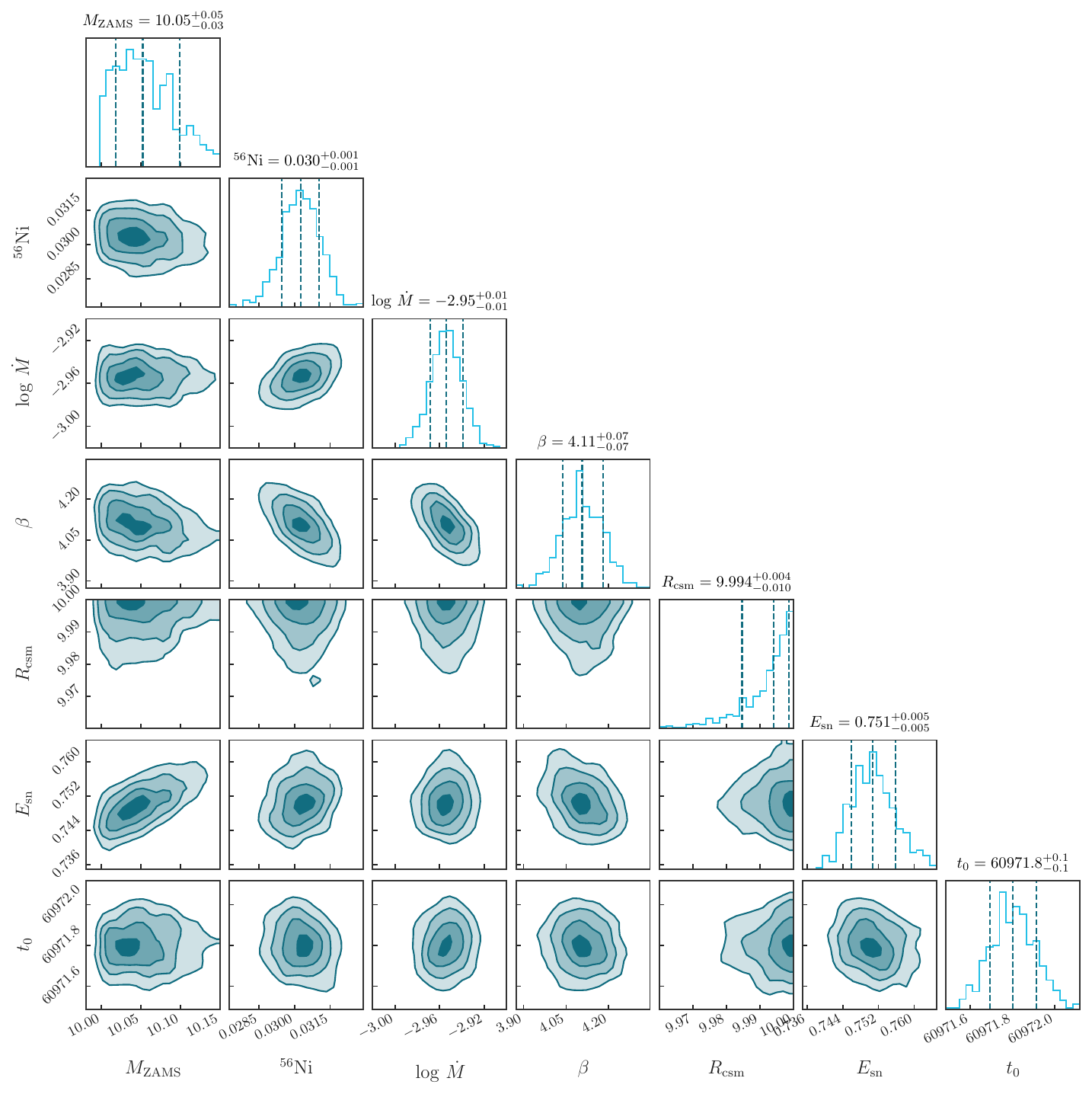}
    \hfill
    \includegraphics[width=0.48\linewidth]{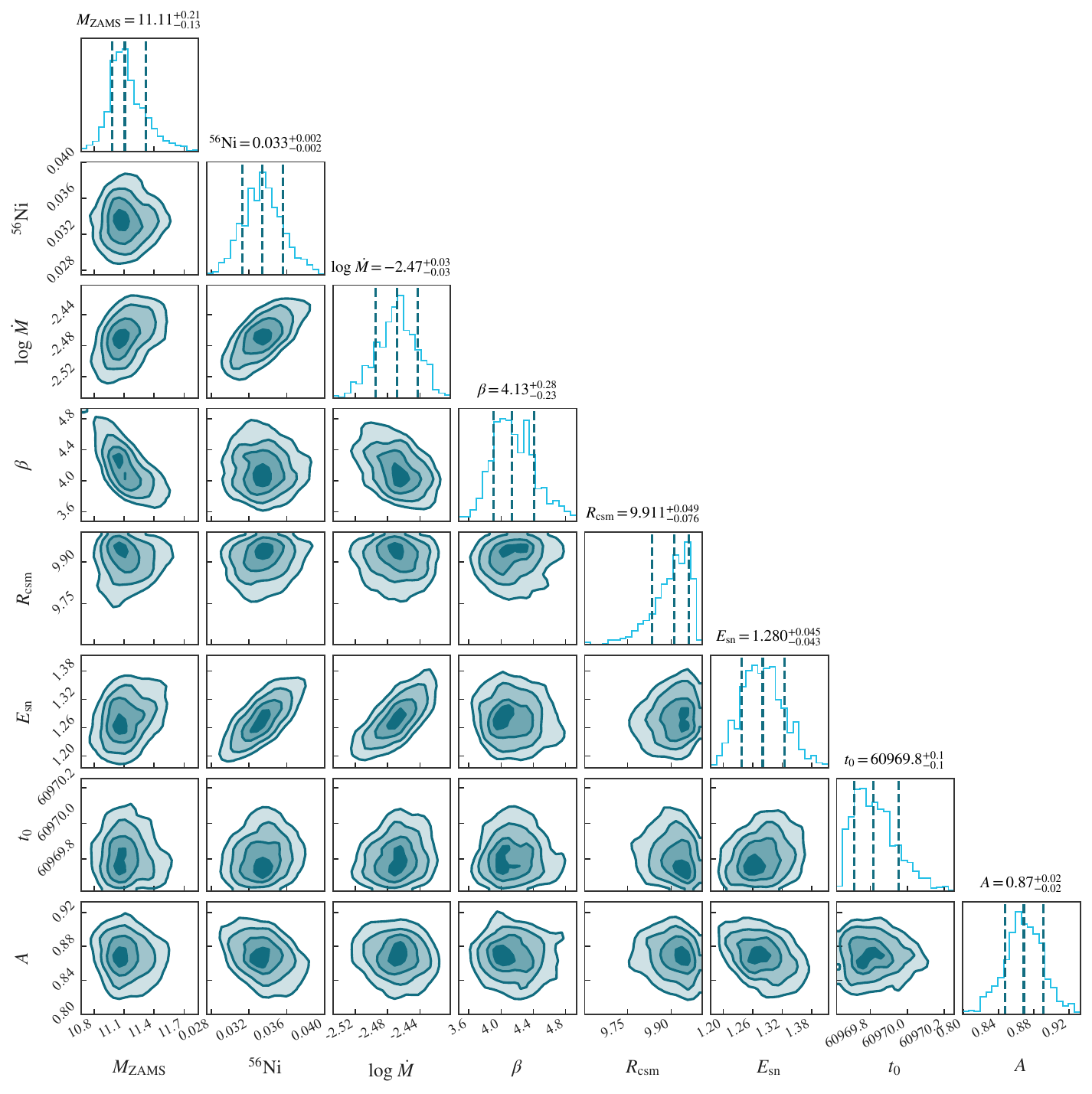}
    \caption{\textit{Left panel:} posterior corner plot for SN\,2025abyc from the surrogate model fit.
             \textit{Right panel:} posterior corner plot from the GPU-accelerated machine-learning reconstruction of the surrogate model.
            The fitted parameters are the ZAMS progenitor mass ($M_{\rm ZAMS}$), synthesised $^{56}$Ni mass ($\rm ^{56}Ni$), mass-loss rate ($\dot{M}$), steepness of the CSM density profile ($\beta$), CSM radius $R_{\mathrm{CSM}}$, SN explosion energy ($E_{\rm sn}$), and explosion epoch ($t_0$).
            The right panel additionally includes a nuisance amplitude parameter $A$ that absorbs global fitting residuals.}
    \label{fig:Corner}
\end{figure*}

\FloatBarrier
\section{Supplementary tables}

\begin{table*}[t]
\centering
\tiny
\setlength{\tabcolsep}{10pt}
\caption{SN\,2025abyc photometry in $ugriz$ bands based on the LJT and MATCH.}
\label{tab:phot_data}
\begin{tabular}{ccccccc}
\hline\hline
MJD & Phase & $u$ & $g$ & $r$ & $i$ & $z$ \\
\hline
60987.7 & 13.1 & -- & 17.31 $\pm$ 0.07 & 17.36 $\pm$ 0.05 & 17.51 $\pm$ 0.06 & 17.57 $\pm$ 0.03 \\
60988.6 & 14.0 & 17.77 $\pm$ 0.01 & 17.31 $\pm$ 0.04 & 17.33 $\pm$ 0.04 & 17.49 $\pm$ 0.04 & 17.57 $\pm$ 0.10 \\
60989.6 & 15.0 & 17.87 $\pm$ 0.01 & 17.32 $\pm$ 0.02 & 17.36 $\pm$ 0.03 & 17.53 $\pm$ 0.06 & 17.56 $\pm$ 0.03 \\
60992.7 & 18.1 & -- & 17.37 $\pm$ 0.05 & 17.36 $\pm$ 0.05 & 17.56 $\pm$ 0.06 & -- \\
60996.6 & 22.0 & 18.70 $\pm$ 0.02 & 17.54 $\pm$ 0.05 & 17.39 $\pm$ 0.04 & 17.53 $\pm$ 0.02 & 17.66 $\pm$ 0.04 \\
60999.7 & 25.1 & 19.01 $\pm$ 0.02 & -- & 17.39 $\pm$ 0.04 & 17.53 $\pm$ 0.06 & 17.64 $\pm$ 0.06 \\
61007.6 & 33.0 & -- & 17.98 $\pm$ 0.04 & 17.49 $\pm$ 0.03 & 17.59 $\pm$ 0.04 & -- \\
61008.6 & 34.0 & 19.71 $\pm$ 0.07 & 18.02 $\pm$ 0.03 & 17.53 $\pm$ 0.04 & 17.59 $\pm$ 0.06 & 17.69 $\pm$ 0.07 \\
61009.6 & 35.0 & 19.85 $\pm$ 0.08 & -- & -- & -- & 17.67 $\pm$ 0.07 \\
61011.6 & 37.0 & -- & 18.10 $\pm$ 0.10 & 17.58 $\pm$ 0.05 & 17.63 $\pm$ 0.07 & -- \\
61012.5 & 37.9 & -- & 18.11 $\pm$ 0.08 & 17.55 $\pm$ 0.23 & 17.72 $\pm$ 0.08 & -- \\
61013.6 & 39.0 & -- & -- & 17.59 $\pm$ 0.07 & 17.69 $\pm$ 0.06 & -- \\
61019.6 & 45.0 & -- & 18.12 $\pm$ 0.09 & 17.61 $\pm$ 0.04 & 17.68 $\pm$ 0.06 & 17.73 $\pm$ 0.12 \\
61029.6 & 55.0 & -- & 18.43 $\pm$ 0.08 & 17.74 $\pm$ 0.04 & 17.81 $\pm$ 0.05 & 17.82 $\pm$ 0.07 \\
61035.6 & 61.0 & -- & 18.57 $\pm$ 0.04 & 17.79 $\pm$ 0.04 & 17.86 $\pm$ 0.05 & 17.77 $\pm$ 0.09 \\
61038.6 & 64.0 & -- & -- & 17.78 $\pm$ 0.02 & 17.90 $\pm$ 0.05 & -- \\
61039.6 & 65.0 & -- & -- & -- & 17.81 $\pm$ 0.12 & -- \\
61042.6 & 68.0 & -- & -- & -- & 17.80 $\pm$ 0.12 & -- \\
61044.6 & 70.0 & -- & -- & -- & 17.93 $\pm$ 0.06 & -- \\
61045.6 & 71.0 & -- & -- & -- & 18.04 $\pm$ 0.12 & -- \\
61049.6 & 75.0 & -- & -- & -- & 18.07 $\pm$ 0.04 & -- \\
61053.5 & 78.9 & -- & 19.03 $\pm$ 0.08 & 18.10 $\pm$ 0.07 & 18.17 $\pm$ 0.05 & 18.10 $\pm$ 0.07 \\
61065.5 & 90.9 & -- & 19.50 $\pm$ 0.15 & 18.57 $\pm$ 0.05 & 18.61 $\pm$ 0.04 & -- \\
61066.5 & 91.9 & -- & -- & -- & 18.76 $\pm$ 0.17 & -- \\
61071.6 & 97.0 & -- & -- & -- & 18.65 $\pm$ 0.15 & -- \\
61074.6 & 100.0 & -- & 20.33 $\pm$ 0.39 & 19.21 $\pm$ 0.08 & 19.24 $\pm$ 0.09 & -- \\
61078.5 & 103.9 & -- & -- & -- & 19.77 $\pm$ 0.22 & -- \\
61095.5 & 120.9 & -- & $<$\,20.88 & 20.15 $\pm$ 0.39 & $<$\,20.58 & -- \\
\hline
\end{tabular}
\end{table*}

\begin{table*}[t]
\caption{Log of spectroscopic observations of SN\,2025abyc based on LJT.}
\label{table:specinfo}
\centering
\tiny                             
\setlength{\tabcolsep}{12pt}
\begin{tabular}{c c c c c c c} 
\hline\hline
Date & MJD & Phase & Telescope/Instrument & Grism+Slit & Airmass & Exp. time(s) \\
\hline
2025-11-08 & 60987.63 & 13.0 & LJT/YFOSC & G3+2.51" & 1.78 & 2100 \\
2025-11-10 & 60989.66 & 15.0 & LJT/YFOSC & G3+2.51" & 1.63 & 700$\times$2 \\
2025-11-13 & 60992.67 & 18.0 & LJT/YFOSC & G3+2.51" & 1.60 & 700$\times$2 \\
2025-11-17 & 60996.64 & 22.0 & LJT/YFOSC & G3+1.81" & 1.63 & 900$\times$2 \\
2025-11-20 & 60999.70 & 25.1 & LJT/YFOSC & G3+1.81" & 1.63 & 900$\times$2 \\
2025-11-29 & 61008.63 & 34.0 & LJT/YFOSC & G3+2.51" & 1.60 & 2400 \\
2025-12-02 & 61011.57 & 36.9 & LJT/YFOSC & G3+2.51" & 1.74 & 2400 \\
2025-12-05 & 61014.59 & 40.0 & LJT/YFOSC & G3+2.51" & 1.64 & 2400 \\
2025-12-20 & 61029.60 & 55.0 & LJT/YFOSC & G3+2.51" & 1.63 & 2400 \\
2025-12-29 & 61038.58 & 64.0 & LJT/YFOSC & G3+2.51" & 1.65 & 2400 \\
2026-01-17 & 61057.54 & 82.9 & LJT/YFOSC & G3+2.51" & 1.70 & 2800 \\
\hline
\multicolumn{7}{l}{{$^*$Phases are relative to the estimated explosion epoch (MJD = $60974.62$).}}
\end{tabular}
\end{table*}

\begin{table*}[t]
    \centering
    \tiny
    \setlength{\tabcolsep}{14pt} 
    \caption{Properties of the comparison SNe II, including normal SNe IIP,  SNe IIL, and short-plateau SNe IIP.}
    \begin{tabular}{cccccc}
        \hline\hline
        SN & Explosion epoch & Redshift & Distance & $E(B-V)_\mathrm{tot}$ & References \\
         & [MJD] & z & [Mpc] & [mag] & \\
        \hline
        1999em & 51475.1$\pm$1.4 & 0.00239 & 8.2$\pm$2.6 & 0.10$\pm$0.05 & 1 \\
        2004et & 53270.0 & 0.000133 &5.9$\pm$0.4 & 0.41$\pm$0.01& 2 \\
        2005cs & 53548.5$\pm$1.0 & 0.00153 & 7.1$\pm$1.2 & 0.05 & 3 \\
        2009N &  54847.6 & 0.0035 & 21.6$\pm$1.1 & 0.13$\pm$0.02 & 4 \\
        2012aw & 56002.1 & 0.002598 &9.9$\pm$0.1 & 0.074$\pm$0.008& 5 \\
        2017eaw & 57886.5$\pm$0.1 & 0.000133 & 6.85$\pm$0.63 & 0.41 & 6 \\
        2018zd & $58178.39^{+0.15}_{-0.50}$ & 0.002979 & $15.0^{+8.9}_{-1.9}$ & 0.17$\pm$0.05 & 7 \\
        2021yja & 59464.4$\pm$0.1& 0.005307 & $23.4^{+5.4}_{-4.4}$ & 0.104 & 8 \\
        \hline
        2013by & 56403.5$\pm$2 & 0.0038 &14.8$\pm$1.0 & - & 9\\
        2013ej & 56497 & 0.002 & 9.7$\pm$0.5 & 0.06 & 10\\
        2014G & 56669.6$\pm$1.7 & 0.0045  & 24.5 & 0.21$\pm$0.11 & 11\\
        2019nyk & 58713.4$\pm$2.0 & 0.021 &  94.7$\pm$7.4 & 0.033$\pm$0.001 & 12 \\
        \hline
        2006Y & 53766.1$\pm$3.4 & 0.0336 & 146 & 0.111 & 13 \\
        2016egz & 57588.2$\pm$ 2.0 & 0.0232 & 100 & 0.014 & 13 \\ 
        2018gj &  58127.3$\pm$1.4& 0.00454 & $19.61\pm1.37$ & 0.08$\pm$0.02 & 14 \\
        2020jfo & 58973.7 & 0.0052 & $14.5$ & 0.065 & 15 \\
        \hline
        2025abyc & 60974.62$\pm$0.02 & 0.021$\pm$0.002 & 82.03$\pm$9.26 & 0.02 & This work \\
        \hline
    \end{tabular}%
    \begin{flushleft}
    \tiny
    References: 
    1. \citet{2002PASP..114...35L,2003MNRAS.338..939E},
    2. \citet{2006MNRAS.372.1315S,2010MNRAS.404..981M},
    3. \citet{2006MNRAS.370.1752P,2009MNRAS.394.2266P},
    4. \citet{2014MNRAS.438..368T},
    5. \citet{2013MNRAS.433.1871B},
    6. \citet{2019ApJ...876...19S,2019ApJ...875..136V},
    7. \citet{2020MNRAS.498...84Z},
    8. \citet{2022ApJ...935...31H},
    9. \citet{2015MNRAS.448.2608V},
    10. \citet{2014MNRAS.438L.101V,2015ApJ...806..160B,2016MNRAS.461.2003Y},
    11.\citet{2016MNRAS.455.2712B,2016MNRAS.462..137T},
    12.\citet{2024A&A...685A..44D},
    13.\citet{2021ApJ...913...55H},
    14.\citet{2023ApJ...954..155T},
    15.\citet{Sollerman2021A&A...655A.105S,2022ApJ...930...34T,Kilpatrick2023MNRAS.524.2161K,Ailawadhi2023MNRAS.519..248A}.
    \end{flushleft}
    \label{tab:SNe_II_info}
\end{table*}


\end{appendix}
\end{document}